\documentclass[12pt,preprint]{aastex}
\singlespace

\providecommand{\mb}{$\Delta m_{15}(B)$}

\providecommand{\mu}{$\Delta m_{15}(U)$}
\providecommand{\mwone}{$\Delta m_{15}(uvw1)$}

\begin{document}
\title{The Absolute Magnitudes of Type I\lowercase{a} Supernovae \\in the Ultraviolet}
\author{Peter~J.~Brown\altaffilmark{1,2}, 
Peter~W. A.~Roming\altaffilmark{1,3},
Peter Milne\altaffilmark{4}, 
Filomena Bufano\altaffilmark{5},\\
Robin Ciardullo\altaffilmark{1}, 
Nancy Elias-Rosa\altaffilmark{6}, 
Alexei V. Filippenko\altaffilmark{7},
Ryan J. Foley\altaffilmark{8,9}, \\
Neil Gehrels\altaffilmark{10}, 
Caryl Gronwall\altaffilmark{1}, 
Malcolm Hicken\altaffilmark{8}, 
Stephen T. Holland\altaffilmark{10,11,12}, \\
Erik A. Hoversten\altaffilmark{1}, 
Stefan Immler\altaffilmark{10,11,13}, 
Robert P. Kirshner\altaffilmark{8},\\
Weidong Li\altaffilmark{7},
Paolo Mazzali\altaffilmark{5,14,15}, 
Mark M. Phillips\altaffilmark{16},
Tyler Pritchard\altaffilmark{1},\\
Martin Still\altaffilmark{17,18}, 
Massimo Turatto\altaffilmark{19}, \&
Daniel Vanden Berk\altaffilmark{20} }
\altaffiltext{1}{Pennsylvania State University,
                 Department of Astronomy \& Astrophysics,
                 University Park, PA 16802.}  
\altaffiltext{2}{Current Address: Department of Physics \& Astronomy,
  University of Utah, 115 South 1400 East \#201, Salt Lake City, UT
  84112-0830; email pbrown@physics.utah.edu.}
\altaffiltext{3}{Current Address: Southwest Research Institute, 
Department of Space Science, 6220 Culebra Rd., San Antonio, TX 78238-5166}
\altaffiltext{4}{University of Arizona, Steward Observatory, Tucson,
  AZ 85719.}
\altaffiltext{5}{ INAF-Osservatorio Astronomico di Padova, Vicolo
  dell'Osservatorio 5, I-35122 Padova, Italy.}
\altaffiltext{6}{Spitzer Science Center, California Institute of
  Technology, 1200 E. California Blvd. Pasadena, CA 91125.}
\altaffiltext{7}{Department of Astronomy, University of California,
  Berkeley, CA 94720-3411.}
\altaffiltext{8}{Harvard-Smithsonian Center for Astrophysics, 60
  Garden Street, Cambridge, MA 02138.}
\altaffiltext{9}{Clay Fellow.}
\altaffiltext{10}{Astrophysics Science Division, NASA Goddard Space Flight Center, Greenbelt, MD 20771.}
\altaffiltext{11}{Center for Research and Exploration in Space Science and Technology, NASA/GSFC}  
\altaffiltext{12}{Universities Space Research Association.}
\altaffiltext{13}{Department of Astronomy, University of Maryland, College Park, MD 20742, USA.}
\altaffiltext{14}{Max-Planck-Institut f\"{u}r Astrophysik,
  Karl-Schwarzschild-Str. 1, D-85748, Garching bei M\"{u}nchen,
  Germany.}
\altaffiltext{15}{Scuola Normale Superiore, Piazza dei Cavalieri, 7 56126 Pisa, Italy.}
\altaffiltext{16}{Las Campanas Observatory, Casilla 601, La Serena,
  Chile.}
\altaffiltext{17}{Mullard Space Science Laboratory, Department of
  Space and Climate Physics, University College London, Holmbury
  St. Mary, Dorking, Surrey, RH5 6NT, UK.}
\altaffiltext{18}{NASA Ames Research Center, Moffett Field, CA 93045.}
\altaffiltext{19}{Osservatorio Astrofisico di Catania, Via S. Sofia
  78, 95123, Catania, Italy.}
\altaffiltext{20}{Saint Vincent College, Latrobe, PA 15650.}
\begin{abstract}
We examine the absolute magnitudes and light-curve shapes of 14 nearby
(redshift $z = 0.004$--0.027) Type Ia supernovae (SNe~Ia) observed in
the ultraviolet (UV) with the {\sl Swift} Ultraviolet/Optical
Telescope.  Colors and absolute magnitudes are calculated using both a
standard Milky Way (MW) extinction law and one for the Large Magellanic
Cloud that has been modified by circumstellar scattering.  We find
very different behavior in the near-UV filters (uvw1$_{rc}$ covering
$\sim 2600$--3300~\AA\ after removing optical light, and $u \approx
3000$--4000~\AA) compared to a mid-UV filter (uvm2 $\approx
2000$--2400~\AA). The uvw1$_{rc}-b$ colors show a scatter of $\sim 0.3$
mag while uvm2$-b$ scatters by nearly 0.9 mag. Similarly, while the
scatter in colors between neighboring filters is small in the optical
and somewhat larger in the near-UV, the large scatter in the
uvm2$-$uvw1 colors implies significantly larger spectral variability
below 2600~\AA. We find that in the near-UV the absolute magnitudes at
peak brightness of normal SNe~Ia in our sample are correlated with the
optical decay rate with a scatter of 0.4 mag, comparable to that found
for the optical in our sample.  However, in the mid-UV the scatter is
larger, $\sim 1$ mag, possibly indicating differences in
metallicity. We find no strong correlation between either the UV
light-curve shapes or the UV colors and the UV absolute
magnitudes. With larger samples, the UV luminosity might be useful as
an additional constraint to help determine distance, extinction, and
metallicity in order to improve the utility of SNe~Ia as standardized
candles.

\end{abstract}


\section{Type I\lowercase{a} Supernovae as Standard Candles\label{IaSC}}

   Type Ia supernovae (SNe~Ia) are among the most luminous of
astrophysical events, making them useful probes of the distant
universe.  SNe~Ia gave the first concrete evidence that the expansion
of the universe is accelerating \citep{Riess_etal_1998,
  Perlmutter_etal_1999}, and they have been used to constrain
cosmological parameters such as $H_0$, $\Omega$, and $\Lambda$
\citep{Barris_etal_2004, Astier_etal_2006, Hicken_etal_2009b,
  Kessler_etal_2009, Riess_etal_2009} and the nature and evolution of
dark energy \citep{Riess_etal_2004a, Riess_etal_2007,
  Wood-Vasey_etal_2007DE, Freedman_etal_2009}.

This is possible because SNe~Ia have a well-established relationship
between their optical peak luminosity and rate of brightness decline,
making them excellent standardizable candles (see
\citealp{Branch_Tammann_1992}, \citealp{Branch_1998},
\citealp{Leibundgut_2001}, and \citealp{Filippenko_2005} for reviews
of the subject).  This standardizing is done by calibrating the peak
luminosity with distance-independent observables such as the
light-curve shape.  One common measure of this shape is the number of
magnitudes the $B$ band declines in the first 15 days after maximum
light, \mb\ 
\citep{Phillips_1993,Hamuy_etal_1996abs,Phillips_etal_1999,Garnavich_etal_2004}.
Another is how much the light curve must be stretched to match a
template \citep{Goldhaber_etal_2001}, or by fitting the light curves to
templates \citep{Riess_etal_1996_mlcs}.  The importance of
distance-independent luminosity indicators has led to the development
of a large number of photometric and spectroscopic methods (see also
\citealp{Nugent_etal_1995,Tripp_1998,Mazzali_etal_1998,Wang_etal_2005color,
  Guy_etal_2005, Guy_etal_2007,Wang_etal_2006,Bongard_etal_2006, Bailey_etal_2009}).

The observed trend that more luminous SNe have broader light curves
will be referred to here generically as the luminosity-width relation.
The underlying cause of this relation is believed to be the amount of
$^{56}$Ni formed in the SN explosion and the resulting change in the
temperature and ionization evolution
\citep{Nugent_etal_1995,Hoflich_etal_1996,Mazzali_etal_2001,Kasen_Woosley_2007}.
The exact shape of the relation is determined by the interplay between
$^{56}$Ni, which contributes to the SN luminosity and affects the
shape of the light curve via its effect on the opacity, and the total
amount of Fe-group elements produced, including stable isotopes which
only affect the opacity \citep{Mazzali_etal_2007}.  There is scatter
about this relation, which may be due to differences in the progenitor
mass or metallicity \citep{Timmes_etal_2003} or asymmetries
\citep{Kasen_etal_2009}.  A few peculiar outliers to the
luminosity-width relation have also been found, including SNe 2000cx
\citep{Li_etal_2001}, 2002cx \citep{Li_etal_2003}, and 2003fg
\citep{Howell_etal_2006}.

More recent studies have expanded the standard-candle utility beyond
optical wavelengths.  \citet{Meikle_2000} and
\citet{Krisciunas_etal_2004} have reported that SNe~Ia might be
standard candles in the near-infrared (IR; $JHK$ bands) at the 0.2 mag
level without a clear dependence on the decay slope.  Recent
observations by \citet{Wood-Vasey_etal_2008} strengthen this claim,
although \citet{Garnavich_etal_2004} and \citet{Krisciunas_etal_2009}
have shown that some SNe that are subluminous in the optical are also
subluminous in the near-IR\null, possibly indicating a continuous or
bimodal distribution of the near-IR absolute magnitudes of rapidly
declining SNe.

At shorter wavelengths, \citet{Jha_etal_2007} has recently shown that
the $U$-band light curves of SNe~Ia are standardizable, similar to the
optical, and can be used to determine the extinction and distances.
The scatter is larger than in the optical $B$ and $V$ bands, and
\citet{Kessler_etal_2009} have found discrepancies in determinations
of cosmological parameters when including rest-frame $u$-band
measurements.  The source of the discrepancy is still not certain,
underlying the importance of understanding SN~Ia behavior in this
regime.  Here we take the next step to shorter wavelengths, reaching
beyond the atmospheric cutoff to compare the absolute magnitudes of
SNe in the UV.

The redshifting of the light from distant SNe leads to two approaches.
One can either follow the well-understood rest-frame optical light by
observing at longer wavelengths, or continue to observe in the optical
and study the evolution of the rest-frame UV\null.  For
moderate-redshift SNe, both approaches can be taken, and the
information can be combined to place strong constraints on the
reddening and distances of the objects.  At higher redshifts, however,
fewer rest-frame optical bands are observable, especially from the
ground.  Thus, rest-frame UV observations will be crucial for future
studies of high-redshift SNe. The utility of the rest-frame UV has
been discussed by \citet{Aldering_etal_2007} and
\citet{Wamsteker_etal_2006}.

Most UV observations of nearby SNe come from the {\it International
  Ultraviolet Explorer (IUE),} which obtained low-resolution UV
spectra for about twelve SNe~Ia near maximum light
\citep{Cappellaro_etal_1995}, and the {\it Hubble Space Telescope
  (HST)}, which has observed about ten more at various epochs
\citep{Wang_etal_2005, Sauer_etal_2008}; see \citet{Panagia_2003} and
\citet{Foley_etal_2008UV} for reviews of UV SN observations. The sample
of rest-frame UV observations is larger at higher redshifts, because
the UV emission appears in the observed optical bands;
\citet{Ellis_etal_2008} and \citet{Foley_etal_2008} find an increased
dispersion in the near-UV region of the spectrum.

Tests for evolutionary effects by comparing the UV region of low- and
high-redshift SNe is limited by the paucity of low-redshift UV
data. But with its fast turn-around time for observations, flexible
scheduling, and UV capability, the Ultraviolet Optical Telescope
(UVOT; \citealp{Roming_etal_2005}) on board the {\sl Swift} spacecraft
\citep{Gehrels_etal_2004} is well suited to improving this situation
by probing the UV photometric behavior of SNe in greater detail than
ever before \citep{Brown_etal_2009}.  In addition, {\it Swift} has
been used to monitor spectra in the UV, and the satellite has obtained
the best UV spectral sequence of a SN~Ia to date (SN~2005cf;
\citealp{Bufano_etal_2009}).

 
\section{Analysis} 
 
In order to determine whether SNe~Ia are standard candles in the UV,
we must be able to compare their absolute magnitudes.  These are
derived through the equation
\begin{displaymath} 
M_X = m_X - rc_X - {\rm K}_X - A_{{\rm MW},X}- A_{{\rm host},X} - \mu,
\end{displaymath} 
\noindent

where $M_X$ is the absolute magnitude of the SN viewed through filter
$X$, $m_X$ is the apparent magnitude at maximum light in filter $X$,
$rc_X$ corrects for the red tails of the uvw2 and uvw1 filters, K$_X$
is the K-correction, $A_{{\rm MW},X}$ and $A_{{\rm host},X}$ are
(respectively) the foreground (Milky Way) and host-galaxy extinction,
and $\mu$ is the distance modulus.  Each term, of course, has its own
uncertainty, which must be added in quadrature to obtain the formal
uncertainty in $M_X$.

\subsection{Peak Apparent Magnitudes} 

This study includes the 14 {\it Swift} SNe~Ia\footnote{SN~2007cv, which 
is included here, has not previously been spectroscopically classified as 
a Ia.  A UVOT spectrum
  of SN~2007cv showed it to be similar to that of SN~Ia 2005cf (Bufano
  2007, private communication).} from \citet{Brown_etal_2009} and
\citet{Milne_etal_2010} (hereafter designated M10) with low extinction
($E(B-V) < 0.5$ mag) and well-sampled light curves in at least one UV
filter.  This includes two SN~1991T-like SNe (SNe 2007S and 2007cq;
X. Wang, 2009, private communication) and two rapid decliners SN (2005ke
is spectroscopically similar to the prototypical SN~1991bg, while
spectra of SN 2007on are more similar to SNe with a smaller \mb\ of 
$\sim 1.6$; M. Phillips, 2009, private communication).  SN~2005hk, a member
of the SN 2002cx-like subclass, was excluded from this study.  Some
UVOT filter characteristics are listed in Table
\ref{table_uvotfilters}.  Observations of each object began at or
before maximum light in the optical and continued for at least two
weeks.  The underlying galaxy light has been subtracted from the SN
light \citep{Brown_etal_2009} and the magnitudes have been calibrated
to the Vega system \citep{Poole_etal_2008}.  Detailed analysis of the
light-curve shapes and colors will be presented by M10; here we focus
on the peak magnitudes derived therein.

The apparent magnitudes at maximum ($m_X$) for our sample were derived
by M10 from fitting the data near peak brightness with a Gaussian rise
and decay, or, for less well-observed SNe, fitting a mean template
light curve for that filter derived from other SNe~Ia.  More details
on the generation of the UVOT templates and the fitting techniques are
provided by M10. The peak magnitudes in the three UV filters and $ubv$
are given in Table \ref{table_appmags}.  To improve our estimates of
the SN peak magnitudes, we have supplemented our UVOT observations
with some ground-based data.  Specifically, we have combined Johnson
$B$ and $V$ observations with UVOT $b$ and $v$ measurements; these two
systems are similar enough that for the purposes of this study the
data are interchangeable, even without color terms or
S-corrections\footnote{S-corrections account for differences in filter
  shapes \citep{Stritzinger_etal_2002}.  \citet{Wang_etal_2009_2005cf}
  show the S-corrections for SN~2005cf from UVOT $bv$ to $BV$ to be
  $\lesssim 0.02$ mag near maximum light.}
\citep{Li_etal_2006_UVOT,Poole_etal_2008}.  To avoid confusion,
however, we treat UVOT $u$ as distinct from Johnson $U$.  Although we
expect that the trends seen in $u$ will be similar to those found for
$U$ (see \citealp{Jha_etal_2007}), the bluer bandpass of UVOT $u$ makes
transformations difficult, and the data may not be directly
comparable.  The values of \mb, the amount in magnitudes that the
$B$-band curve declines in the 15 days after maximum light, from UVOT
or the literature, are also given in Table \ref{table_appmags}.

\subsection{Red Tail Correction\label{redleak}} 

\begin{figure} 
\resizebox{14cm}{!}{\includegraphics*{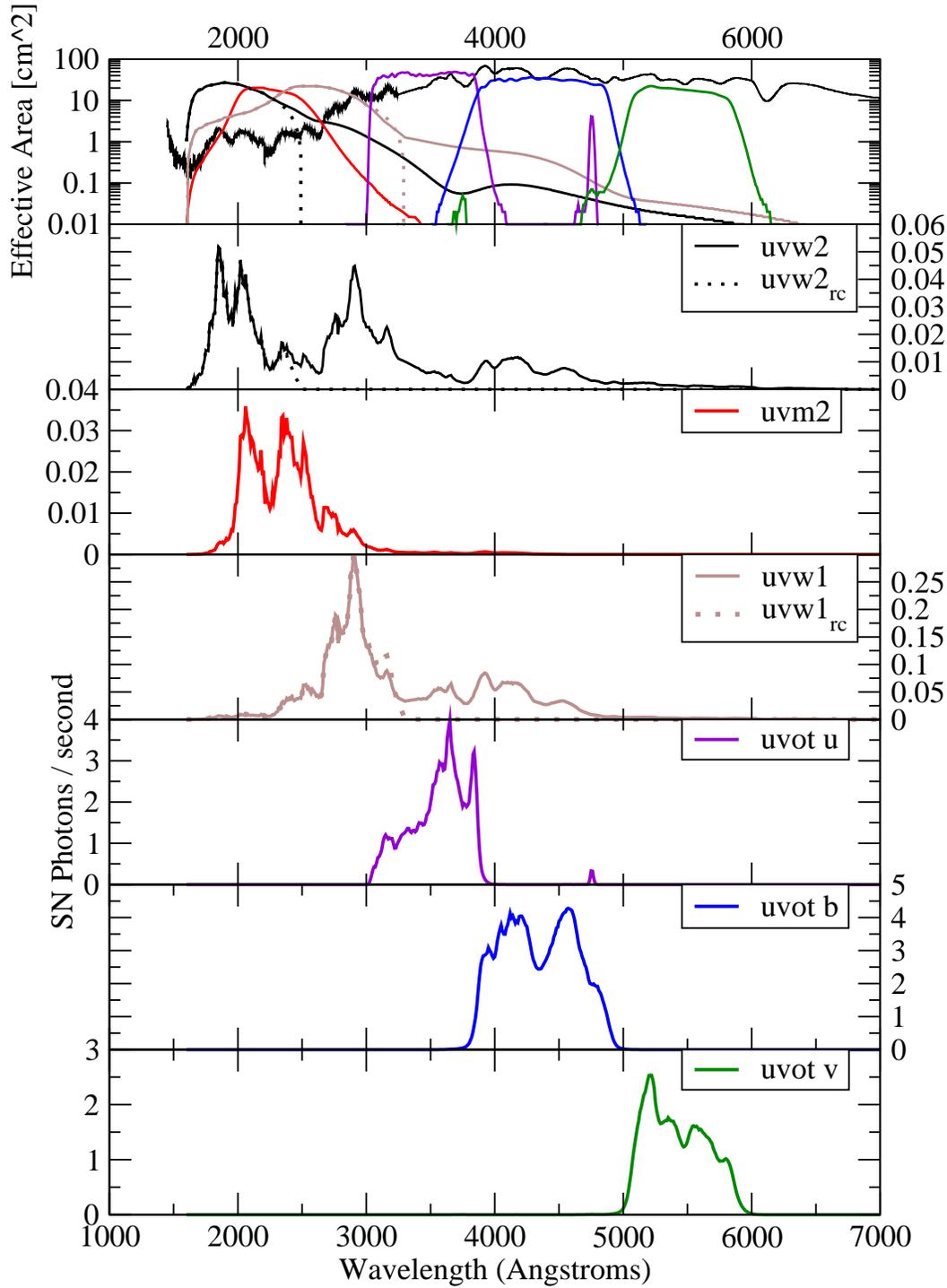}  } 
\caption[UVOT filter curves and the SN~1992A spectrum.] 
{In the top panel, the {\it HST} spectrum of SN~1992A
  \citep{Kirshner_etal_1993} at $\sim 5$ days after maximum brightness
  is plotted with respect to the six UVOT filters on a logarithmic
  scale to reveal the structure in the low-flux UV region.  The six
  lower panels show the distribution of photons resulting from the
  SN~1992A spectrum passing through the six UVOT filters.  }
\label{plot_uvotfilters}  
\end{figure}

While the UV filters have the bulk of their transmission in the
near-UV, the red tails of uvw1 and uvw2 combined with the very red
SN~Ia spectral shape result in a redder distribution of received
photons compared to the transmission curves.  The uvm2 filter does not
have this problem, since its transmission curve is well blocked at
longer wavelengths.  To visualize the photons UVOT should detect from
a SN~Ia near maximum light, we multiplied the SN~1992A UV-optical
spectrum taken at 5 days after maximum light with {\it HST}
\citep{Kirshner_etal_1993} with the effective area curves of the UVOT
filters \citep{Poole_etal_2008}.  The resulting photon distributions
are displayed in Figure \ref{plot_uvotfilters}.  The photon-weighted
effective wavelengths (where an equal number of observed photons are
longer and shorter than the given wavelength) are listed in Table
\ref{table_uvotfilters}.


In order to better understand the intrinsic UV flux and for comparison
with high-redshift observations, it is desirable to estimate the
amount of flux coming from the tails and subtract it off.  To do this,
we first define new filter curves, which we designate uvw2$_{rc}$ and
uvw1$_{rc}$ (for ``red corrected'').  These take the latest
calibration of the effective areas of the uvw2 and uvw1 filters
(Landsman, CALDB document, in preparation).  From the uvw2 effective
area at 2200~\AA\ (3000~\AA\ for uvw1$_{rc}$), the new curves decrease
linearly to zero at 2500~\AA\ (3300~\AA\ for uvw1$_{rc}$).  These
curves, and the corresponding photon distribution for the SN~1992A
spectrum, are displayed as dashed lines in Figure
\ref{plot_uvotfilters}.  New zeropoints are determined by passing the
spectrum of Vega through the filters and setting its magnitude to
zero, thus placing these magnitudes on the Vega system:
$ZP_{\rm w2rc} = 17.25$ and $ZP_{\rm w1rc} = 17.34$ mag, compared to
$ZP_{\rm uvw2} = 17.35$ and $ZP_{\rm uvw1} = 17.49$ mag \citep{Poole_etal_2008}.
The UVOT magnitudes are determined by $m_X= -2.5 {\rm log} (\int
\lambda/hc f_{\lambda} E_{X,\lambda} d\lambda) + {\rm Zpt}_X$.
For each SN, we started with one of three spectra: the {\it HST}
spectrum of SN~1992A from \citet{Kirshner_etal_1993}, a maximum-light
template from \citet{Nugent_etal_2002}, which incorporated additional
{\it IUE} UV spectra, or a maximum-light template from
\citet{Hsiao_etal_2007}, which added more {\it HST} UV spectra and
rest-frame UV spectra from higher redshift SNe.  The input spectrum
was shifted to the rest frame of the SN and multiplied by a spline
function passing through the Vega effective wavelengths of the UVOT
filters (uvw2 excluded here) and the ratio of the synthetic to the
observed counts in each filter.  The uvw2 and $v$ ratios were also used
to anchor the spline at the end points of the spectrum, 1600~\AA\ and 8000
\AA, respectively.  Such a color correction to the spectrum is referred
to as ``mangling'' \citep{Hsiao_etal_2007}.  This procedure was
repeated until the count rates from the mangled spectrum matched the
photometry.  

To account for the red tails, after the first iteration
the uvw2 and uvw1 ratios were scaled by the fraction of the counts in
the UV portion (from the previous iteration) in order to better
approximate the flux at the reference wavelength.  The uvw2 ratio was
set to zero when the counts through the tail of the filter already
accounted for all of the observed counts.  The spectrum was passed
through the original and ``red corrected'' filters (e.g., uvw2 and
uvw2$_{rc}$), and the magnitude difference was used to transform the
observed magnitudes to red tail corrected magnitudes such that
uvw2$_{rc} = $ uvw2$-$rc$_{w2}$.  This was repeated twenty times for each
of the three template spectra, each time randomly offsetting the peak
magnitudes consistent with the photometric errors and repeating the
mangling and synthetic photometry.  The mean and standard deviation
were taken to be the correction and its corresponding error.  This
procedure is not unlike the S-corrections used to correct observations
to a standard filter set from filters covering a similar wavelength
range but with transmission varying from the standard filter set.  In
this case the two filters are by definition identical everywhere
except at the long-wavelength tail.

The fraction of the estimated counts in each ``red-corrected filter''
to that of the original filter, and the correction in magnitudes, is
listed in Table \ref{table_redtail}.  For uvw1$_{rc}$ this fraction
varies from 0.68 down to 0.29, corresponding to a change in magnitudes
(including the different zeropoint) of 0.22 to 0.82 mag.  For
uvw2$_{rc}$ the largest fraction is 0.45, corresponding to a magnitude
difference of 0.63.  Several have essentially no flux in the
uvw2$_{rc}$ filter, with the magnitude of the fraction and correction
being largely dependent on how much the iterative mangling depressed
the UV portion before the iterations stopped.  While this could
signify extreme variations in the UV flux, it is more likely
indicative of less extreme, but nonetheless very significant, spectral
variations in the UV that are not well replicated by mangling the
available spectral templates.  The shape of these variations can be
explored through theoretical modeling
\citep{Hoflich_etal_1996,Lentz_etal_2000,Sauer_etal_2008} or through
additional UV spectra, such as those obtainable with the Cosmic Origins
Spectrograph (COS) on the refurbished {\it HST}.  Recent theoretical
modeling does indicate that changes in metallicity have the largest
effect in this wavelength region \citep{Sauer_etal_2008}, and
comparisons with the growing {\it Swift} SN sample should put
observational constraints on the metallicity range of the observed
SNe.  Since the correction factors for uvw2$_{rc}$ are large and
uncertain, we will exclude them from our analysis of absolute
magnitudes.


For comparison with our SN~Ia corrections, and for more general use,
similar correction factors were also computed for a large sample of
spectra, including stars, galaxies, blackbodies, and other SNe.  
These are detailed in Appendix A.  Figure \ref{plot_redcor} shows the
correction factors as a function of the UV$-b$ colors along with the
sample of SNe~Ia studied here.  The correction factors for a subsample
of these are tabulated in Appendix A.  Especially for uvw2, the
red-tail correction is not a monotonic function of a single color, but
different sources of the same color have corrections differing by up
to half a magnitude at uvw2$-b \approx 2$.  Beyond that, the
corrections are too steep to be reliable, but the UV would account for
less than 10\% of the detected counts anyway.  For objects
whose general classification is known, the red correction can be
determined using the observed UV magnitude and an optical magnitude
(which could be obtained from the ground).  Otherwise, multiple colors
are likely necessary to accurately determine the correction factors in
the absence of a suitable template spectrum.

Returning to the SN corrections, the SN~Ia colors and calculated
corrections place them roughly on the track of the blackbody and
stellar spectra.  The uvw2$_{rc}$ corrections are also broadly
consistent with low-temperature blackbodies and late-type stellar
spectra, but on the steep dropoff where small changes in color
correspond to large corrections.  This makes the corrections highly
uncertain, as differences in the UV spectral shape or errors in the
photometry can easily result in an incorrect estimate of the UV flux.
Thus, the uvw2 filter will be excluded from this analysis pending
further study.

\subsection{K-corrections} 

Despite the low redshift range of our sample, K-corrections
\citep{Oke_Sandage_1968} are non-negligible due to the sharp drop in the
SN flux at shorter wavelengths.  Thus, even a small shift of the
spectrum to longer wavelengths results in less flux being transmitted
through the UV filters.  To estimate the effect of the K-correction,
we have used the template UV spectra mangled to match the observed
photometry as described in \S \ref{redleak}.  The mangled spectrum was
then deredshifted by the host-galaxy redshift, and the synthetic
photometry was repeated.  The K-correction was then the difference
between the observed magnitude of the redshifted spectrum and the
observed magnitude of the rest-frame spectrum plus the flux-dilution
factor $2.5 {\rm log}(1+z)$ \citep{Oke_Sandage_1968}.  

The derived K-corrections for each SN are listed in Table
\ref{table_kcor}.  The K-corrections in $b$ and $v$ are small, as
previously determined by \citet{Hamuy_etal_1993,Nugent_etal_2002}, but
are larger in the UV (see \citealp{Jha_etal_2006U} for a discussion of
ground-based $U$-band K-corrections).  Uncertainties were estimated
from the difference in the K-corrections after offsetting the observed
photometry by the photometric errors and repeating the mangling and
determination of K-corrections.

More multi-epoch UV spectra of a range of SNe~Ia are needed to better
understand the UV photometry (including K-corrections) and how much
the spectra vary with time and among objects.  Spectral differences
between SNe may have different effects than accounted for by the
mangling, particularly for the SN~1991T and SN~1991bg subtypes.  The
bulk differences, however, are taken care of by the mangling, and the
differences in the red-tail and K-corrections between the SN~1992A
spectrum, the \citet{Hsiao_etal_2007} and \citet{Nugent_etal_2002}
templates, and even a flat spectrum after mangling are small in the
filters considered.  Only uvw2 shows significant differences due to
the lack of a reliable flux point at shorter wavelengths.


\subsection{Correcting for Extinction} 

The extinction to a SN has multiple components, as light from the SN
can be extinguished by dust surrounding the progenitor, in the host
galaxy, in between galaxies, and in the Milky Way.  The latter is easy
to estimate: we can simply use the reddening maps of
\citet{Schlegel_etal_1998}, and apply a \citet{Cardelli_etal_1989}
extinction law with $R_V=3.1$, corresponding to the average value in
the MW.
This provides a lower limit to the amount of attenuation undergone by
the SN.  Estimating the amount of extinction from the other sources,
and thus the total line-of-sight extinction, is more difficult.

To estimate the total reddening, we can take advantage of the fact
that most SNe~Ia have similar post-maximum color evolution in the
optical and peak optical colors related to their \mb\ value.  The
observed colors can therefore be compared to the expected colors to
produce an estimate of the reddening.  Table \ref{table_EBV} lists
three different estimates of this reddening, one based on the color at
peak brightness \citep{Phillips_etal_1999,Garnavich_etal_2004}, one
using the supernova color 30 to 60 days after maximum light
\citep{Lira_1995, Phillips_etal_1999}, and two derived from the
updated version of the multi-color light-curve shape method (MLCS2k2;
\citealp{Riess_etal_1996_mlcs,Jha_etal_2007,Hicken_etal_2009a}).  For
MLCS2k2 we use the values of \citet{Hicken_etal_2009a} derived using
the MW value of $R_V = 3.1$ as well as the $R_V = 1.7$ value which
they find gives the best results; these are referred to hereafter as
MLCS31 and MLCS17, respectively.  In the final analysis, we first
adopt the appropriate MLCS2k2 reddening if available, because it
considers the colors at all epochs.  Otherwise, we use the average of
the reddening values determined from the tail and peak colors if
available, or just the peak reddening.  From the total reddening
toward the SN determined by the SN colors with the above methods, the
contribution from the MW along the line of sight in the
\citet{Schlegel_etal_1998} reddening maps is subtracted, and the rest
is considered together as the host-galaxy reddening.  
The reddening from the MW and host-galaxy (determined with different methods) 
are listed in Table 5.

The \citet{Phillips_etal_1999} color relation only applies to the
range $0.9 < $\mb$ < 1.6$ mag, so for the rapidly declining SNe~2005ke and
2007on (\mb\ equal to 1.77 and 1.89 mag, respectively) we use the
relationship of \citet{Garnavich_etal_2004}.  Both of these SNe,
however, appear to have peculiar colors.  SN~2005ke is very red at
peak, which could be interpreted as a significant reddening of $E(B-V)
\approx 0.4$ mag, but the later tail colors and MLCS2k2 fitting (see
below) imply a very small host-galaxy reddening, and no interstellar
absorption from the host is detectable in high-resolution spectra
(Foley 2008, private communication).  SN~2007on appears very blue at
peak, which results in a negative extinction correction when compared
with the peak or tail colors for SNe of similar decay rates; hence, we
set the host reddening to be equal to zero.  The optical decay rate is
also affected by reddening \citep{Phillips_etal_1999}, so wherever we
use \mb, the corrected decay rate \mb$_{true}$ is implied (though this
affects the SNe in our sample by at most 0.05 mag).

\begin{figure} 
\resizebox{14cm}{!}{\includegraphics*{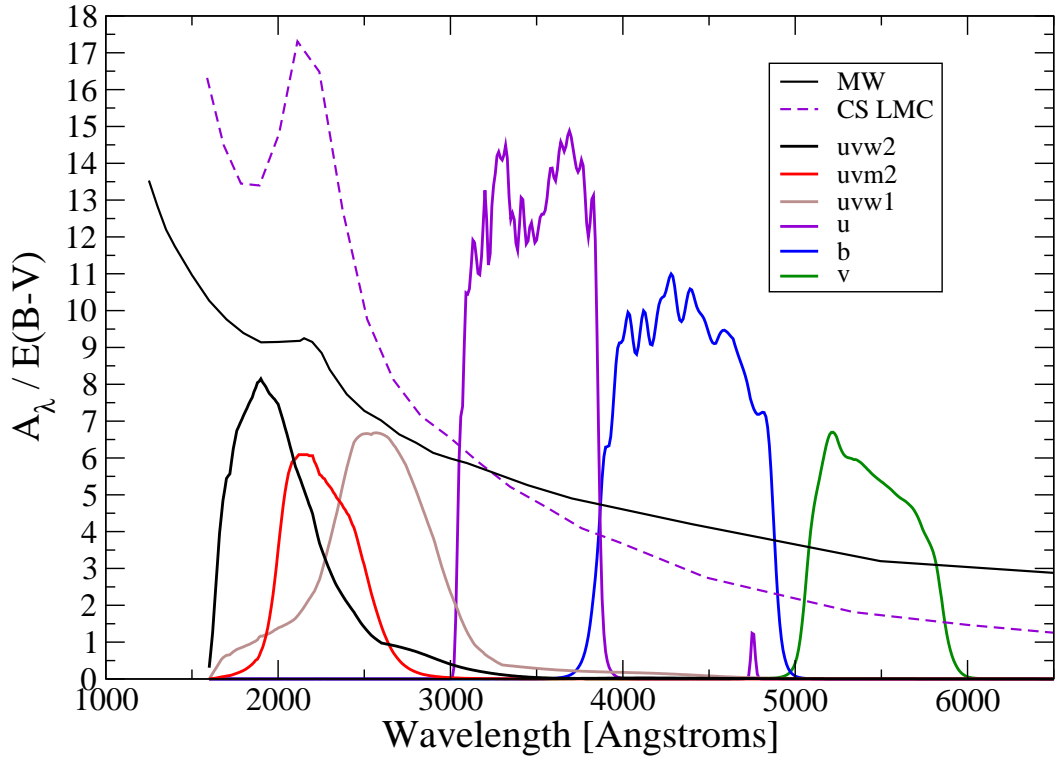}  } 
\caption[UVOT filter curves and reddening laws.]  {The ratio of
  selective to total extinction for the MW \citep{Cardelli_etal_1989}
  and CSLMC \citep{Goobar_2008} extinction laws plotted as a function
  of wavelength with the UVOT effective area curves.  }
\label{plot_excurves}  
\end{figure}

The next step is to convert from the $E(B-V)$ reddening to the
extinction in each filter.  The wavelength dependence of the
extinction curves toward SNe is not completely understood, though
several well-studied, highly extinguished SNe~Ia have been found to
have extinction differing from that in the MW (e.g.,
\citealp{Elias-Rosa_etal_2006, Wang_etal_2008}), and many samples have
shown SNe to have systematically lower values of $R_V$ than the
canonical MW value of 3.1
\citep[e.g.,][]{Riess_etal_1996_dust,Jha_etal_2007,Hicken_etal_2009b,
  Nobili_Goobar_2008,Kessler_etal_2009,Folatelli_etal_2010}.  Whether
the UV extinction follows the \citet{Cardelli_etal_1989}
parameterization according to $R_V$ is also uncertain, and the
intrinsic reddening law of circumstellar material might be modified by
geometric effects \citep{Wang_2005,Goobar_2008}.  For this study we
will calculate the host extinction using the $R_V=3.1$ extinction law
of \citet{Cardelli_etal_1989} corresponding to that of the MW (and
hereafter designated as MW), and the circumstellar LMC law of
\citet{Goobar_2008} (hereafter referred to as CSLMC)extended into the
UV (Goobar 2010, private communication).  The ratios of total to
selective extinction for these curves are plotted in Figure
\ref{plot_excurves}.  In particular, we point out that while the CSLMC
ratio is smaller in the optical, consistent with that found in
optical studies, the curves cross over in the $u$ band and are much
larger in the UV.  The true extinction may not correspond to either of
these laws, as grain size and distribution may effect the strength of
the mid-UV bump and the far-UV rise, and the total extinction may be
the sum of components with differing wavelength dependence.  It is
clear, however, that the UV region is a very sensitive probe of these
effects, and future work will include folding these UV data into a
multi-wavelength study of extinction.  M10 explores a color relation
that might be employed to study UV extinction.

To study the effect of extinction in each filter for a SN spectrum, we
took the {\it HST} UV-optical spectrum of SN 1992A
\citep{Kirshner_etal_1993} and extinguished it with the MW extinction
law and differing amounts of $E(B-V)$ reddening.  We calculated the
effective extinction in each UVOT filter by comparing synthetic
photometry in the reddened and unreddened spectra.  The same was done
with the \citet{Goobar_2008} CSLMC extinction law.  This method is
preferred to using a single wavelength (e.g., central, peak, or
effective wavelength) to represent the entire filter, since that
wavelength is not necessarily representative of the integrated photon
distribution.

\begin{figure} 
\plottwo{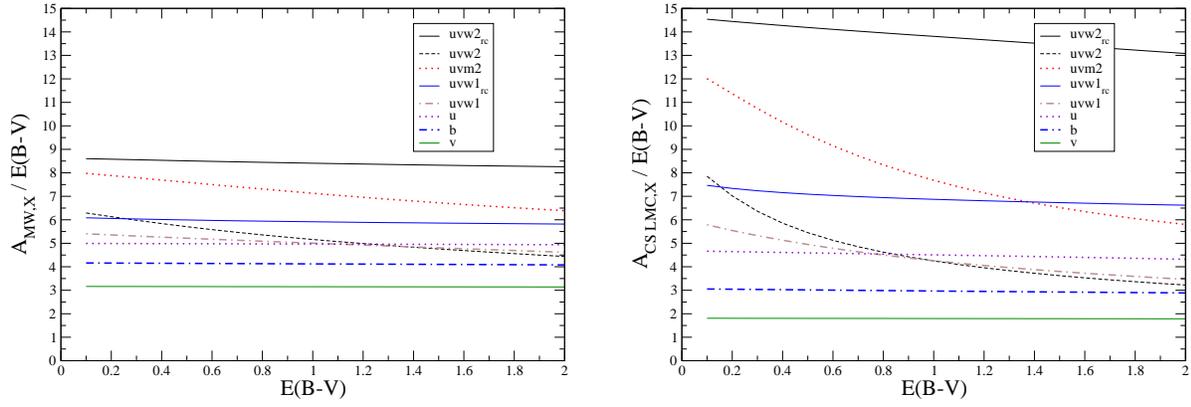}{f4b_color.eps}
\caption[$A_X$ divided by $E(B-V)$ as a function of reddening using
  the MW and CSLMC extinction laws.]  {$A_X$ divided by $E(B-V)$ as a
  function of $E(B-V)$ for the six UVOT filters.  The MW values are
  plotted in the left panel and the CSLMC in the right. }\label{plot_rv}
\end{figure}

For a general understanding of the behavior of the extinction laws, we
followed the above steps for a range of $E(B-V)$ values and fit it
with an analytic function.  Usually, the total extinction through a
given filter, $A_X$, is related to the differential extinction in a
linear fashion, $A_X = R_X E(B-V)$.  Indeed, this is the case for the
UVOT $u$, $b$, and $v$ filters, all the way through $E(B-V) = 2$ mag.
However, as Figure \ref{plot_rv} illustrates by plotting $A_X/E(B-V)$,
the relationship for the UVOT UV filters is highly nonlinear.  For
both extinction curves, the coefficients for a quadratic fit for
$A_\lambda$ in terms of $E(B-V)$, such that $A_\lambda = R_{1,\lambda}
E(B-V) + R_{2,\lambda} E(B-V)^2 $, are given in Table
\ref{table_uvotfilters}.  As evidenced by those coefficients and
Figure \ref{plot_rv}, the quadratic term is unnecessary in the optical
but important in the UV.  This can be understood in terms of the flux
distribution: because dust preferentially extinguishes bluer photons,
the effective wavelength of the UVOT broad-band filters becomes redder
as the extinction increases. However, this dependence on the spectral
shape also means that the extinction coefficients vary with the
significantly different colors of the SNe. 

To account for both of these effects, extinction coefficients were
determined for each SN individually as part of the mangling described
above.  The observer-frame mangled spectrum was first dereddened by
the $R_V=3.1$ MW law based on the \citet{Schlegel_etal_1998}
reddening, and extinction coefficients were determined such that
$R_X = A_X/E(B-V)$.  While the effect of MW reddening will change with
redshift, in particular as the 2200~\AA\ bump moves in relation to the
broad emission and absorption features in the UV spectrum of the SN,
the differences are dominated by the varying spectral shapes of the
SNe.
These are tabulated in Table 6.  $R_X$ values for a variety of sources are tabulated in Appendix B, 
and the coefficients for the SNe are consistent with those of other
sources with similar colors.  After this the spectrum was deredshifted
to the SN host frame, and dereddened based on the estimated host
$E(B-V)$ together with the $R_V = 3.1$ MW law and the CSLMC law, and
host extinction coefficients determined as above.

\subsection{Determining the Distances} 

Five of our SN host galaxies have independent distance estimates or
are in groups whose distances are known from various techniques.  NGC
524 (host galaxy of SN 2008Q), NGC 1404 (host galaxy of SN 2007on),
the Fornax Cluster (host to SN 2005ke in NGC~1371), and the Hydra
Cluster (host to SN 2007cv in NGC~3311) all have distances from the
surface brightness fluctuation (SBF) method \citep{Jensen_etal_2003,
  Tonry_etal_2001, Mieske_etal_2005}.  These were all placed on the
same absolute scale (based on the \citep{Freedman_etal_2001}
measurement of the Hubble constant) by adjusting the
\citet{Tonry_etal_2001} measurements by 0.16~mag, to match the
IR-based zeropoint of \citet{Jensen_etal_2003}.  Additionally, SN
2005am (in NGC 2811) and SN 2005df (in NGC~1559) have distance moduli
from the luminosity vs. line width (Tully-Fisher) relation, though we
do not use the measurement for NGC 2811 because of its large (0.8 mag)
uncertainty \citep{Tully_etal_1992}.  The adopted distances are listed
in Table \ref{table_host}.

For the remaining SNe, we adopted a distance based on the recession
velocities of their host galaxies, a model for the local velocity flow
\citep{Mould_etal_2000}, and $H_0 = 72 $ km~s$^{-1}$~Mpc$^{-1}$
\citep{Freedman_etal_2001}.  We ignore the stated 8
km~s$^{-1}$~Mpc$^{-1}$ uncertainty in the Hubble constant since an
error would cause a shift in the absolute magnitudes but not affect
the scatter, which is our primary measure of the standardizability of
SNe~Ia in the UV.  The Hubble-flow distances for all of the SNe are
listed in Table~\ref{table_host}, along with an uncertainty which is
based on a possible peculiar motion of 150~km~s$^{-1}$, but in the
subsequent analysis we use only the Hubble-flow distances for the SNe
which do not have an independent distance measurement.

Finally, we check our distance estimates by exploiting the fact that
SNe~Ia are standardizable candles in the optical.   Table
\ref{table_host} gives the distance measured using MLCS2k2 with
both $R_V = 3.1$ and 1.7 (scaled to the same H$_0=72$ scale used above.
 Eight of these SN distances were reported by
\citet{Hicken_etal_2009a} and an MLCS2k2 distance 
with $R_V = 3.1$ for SN~2005df was reported by M10. Of the
nine SNe with MLCS2k2 distances, six of the SN distances are within
$1\sigma$ of the adopted distance, and three are about $2\sigma$ away
(with the uncertainties added in quadrature).

We also used NED\footnote{http://nedwww.ipac.caltech.edu/.} and
HyperLeda\footnote{http://leda.univ-lyon1.fr.}
\citep{Paturel_etal_2003} to find the Hubble morphological type
(translated from the de Vaucouleurs scale when necessary;
\citealp{devac_etal_1991}).  The classifications and individual
references are given in Table \ref{table_host}. The majority of the
SNe in our sample are found in late-type galaxies.  Of those in
early-type galaxies, the hosts of SNe 2005cf, 2006dm, and 2006ej are
distorted by interactions with other galaxies, so the progenitor could
have arisen from relatively recent star formation.  A larger sample of
SNe arising from older stellar populations is needed to compare UV
properties from different environments.


\section{Results} 
\subsection{Peak Pseudocolors vs. Reddening\label{section_colorsvebv}} 

\begin{figure} 
\resizebox{14cm}{!}{\includegraphics*{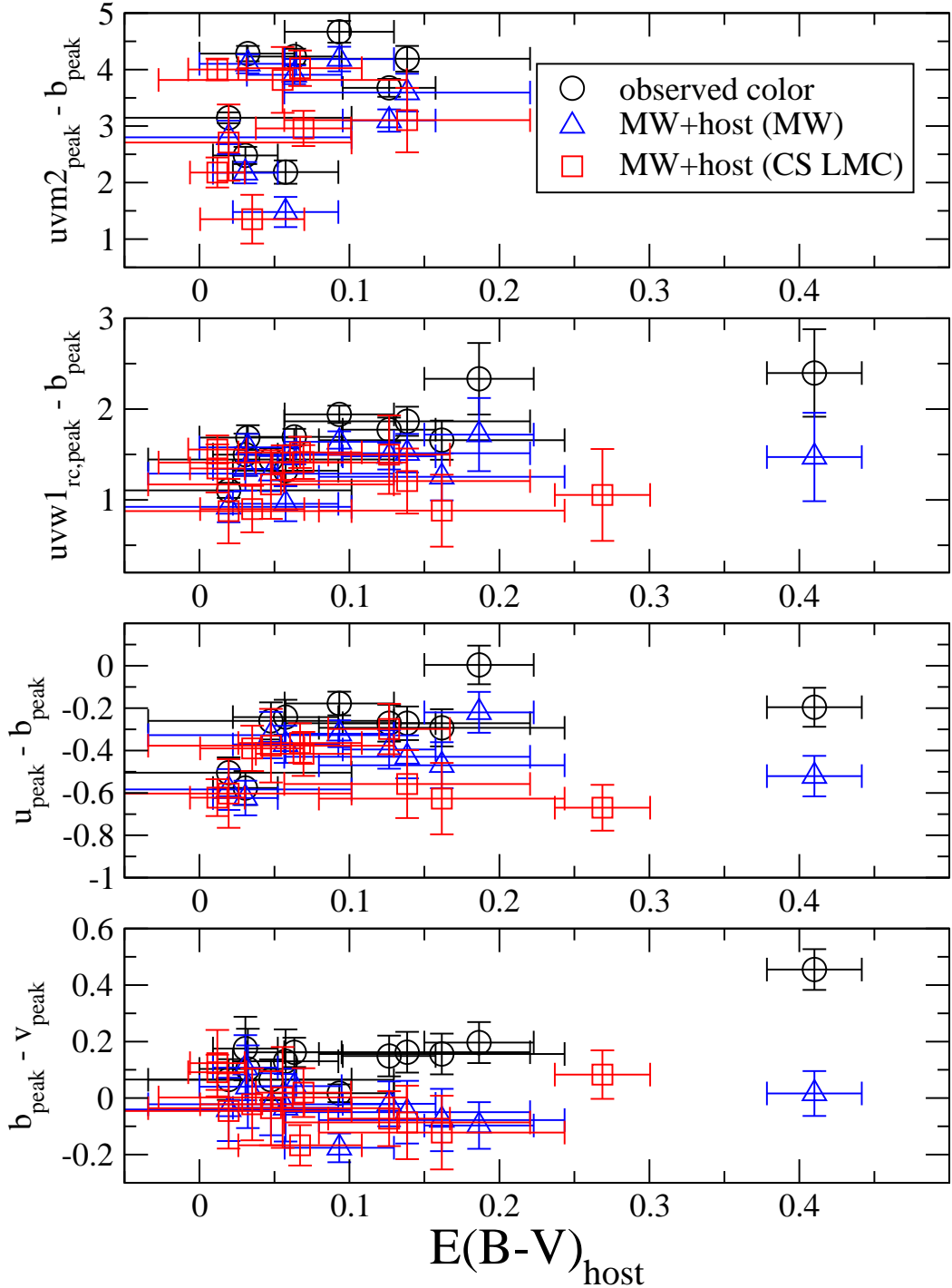}  } 
\caption[UV$-b$ colors of the $\Delta m_{15}(B) < 1.6$ mag SNe~Ia.]
        {UV$-b$ colors of the SNe~Ia with $\Delta m_{15}(B) < 1.6$ mag
          without extinction correction and after correction for
          Galactic and host-galaxy reddening (corrected with the MW
          and CSLMC extinction laws).  For the observed colors, the
          abscissa corresponds to the MLCS31 host reddening.  Since the
          MLCS17 model gives a different value for the host reddening,
          the same SNe do not always line up vertically.
There does not appear to be any significant difference in the
extinction-corrected colors, though the sample is small and the SNe in
our sample with the highest reddening values also have the lowest
values of \mb.  }\label{plot_colorsvebv}
\end{figure}

We first examine the peak colors to verify that our extinction
correction does not leave or create any bias in the colors.  Figure
\ref{plot_colorsvebv} plots the difference between the maximum UV
magnitude of the normal SN and its maximum $b$ magnitude, corrected
for reddening based on the Galactic and adopted host-galaxy reddening
and the MW and CSLMC extinction laws, as a function of the host
reddening.  These ``pseudocolors" are relatively constant with
$R(B-V)$, so no reddening law is clearly better.  The two SNe with the
highest host reddening are also the SNe with the broadest optical
light curves of the sample, and neither are detected in the uvm2, so
it would be difficult to draw strong conclusions based on them.  A
similar exercise done with a larger sample of moderately extinguished
SNe is needed to determine if one of the reddening laws is superior.
Deeper observations in the uvm2 filter are also required to get a
better handle on the UV extinction, where the differences are largest,
without the uncertainties of the red-tail correction.

The uncertainty in the reddening also propagates to a higher
uncertainty in the extinction because of the large extinction
coefficients in the UV.  For example, a modest change in $E(B-V)$ of
0.05 mag (the dispersion in the Lira relation;
\citealp{Phillips_etal_1999}) shifts $A_{m2}$ by $\sim 0.4$--0.7 mag.
However, this large lever arm can provide useful constraints on the
measurement of interstellar reddening, just as expanding the
wavelength range of the measurements in the other direction to the
near-IR has already increased the accuracy of extinction corrections
\citep{Elias-Rosa_etal_2006, Krisciunas_etal_2007,Mandel_etal_2009}.

\subsection{Peak Pseudocolors vs. \mb\label{section_colorsvdmb15} }

\begin{figure} 
\resizebox{10cm}{!}{\includegraphics*{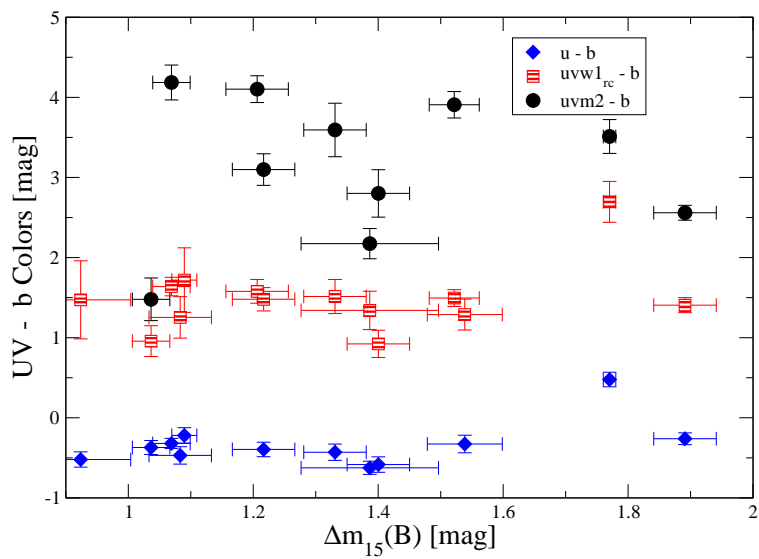}  } 
\caption[Peak UV$-b$ pseudocolors plotted with respect to their
  optical decay rate \mb.]  {Peak UV$-b$ pseudocolors plotted with
  respect to their optical decay rate \mb.  The colors have been
  corrected for Galactic and host reddening using the MW extinction
  law.  SN~2005ke, at $\Delta m_{15}(B) = 1.77$ mag, is much redder
  than the other SNe in $u-b$ and uvw1$_{rc}-b$, but not so different
  in uvm2$-b$. The same behavior is seen when the host reddening is
  corrected with the CSLMC law.   }\label{plot_colorsvdmb15}
\end{figure}

Without the added uncertainties from distances, we can take a first
look at how standard the UV peak magnitudes are by comparing them with
the standardizable optical peak magnitudes.
Figure~\ref{plot_colorsvdmb15} shows the UV-optical pseudocolors of
each supernova at maximum, plotted against optical decay rate, $\Delta
m_{15}(B)$.  The colors are relatively flat with respect to $\Delta
m_{15}(B)$.  SN 2005ke, at $\Delta m_{15}(B) = 1.77$ mag, appears as a
red outlier in $u-b$ and uvw1$_{rc}-b$, but not in uvm2$-b$.  Note
that our $u-b$ plot does not show the reddening trend present in
ground-based $U-B$ data \citep{Jha_etal_2006U}.  However, considering
that the UVOT $u$-band filter is bluer than Johnson $U$, and that the
reddening trend seen by \citet{Jha_etal_2006U} reverses at shorter
wavelengths \citep{Foley_etal_2008UV}, this result is reasonable.

Average reddening-corrected colors for our SNe with $\Delta m_{15}(B)
< 1.6$ mag are listed in Table~\ref{table_colors}.  The colors
corrected with the CSLMC law are up to 0.15 mag bluer due to the
steeper correction, but the differences in the scatter are 0.03 mag or
less.  In the following we compare the scatter in different colors
using the MW-corrected values but the trend is the same for the CSLMC
law.  Compared to the 0.07 mag scatter in the $b-v$ colors, the
scatter increases modestly in the $u-b$ and uvw1$_{rc}-b$ bands to
0.13 and 0.25 mag respectively.  The uvm2$-b$ colors scatter by 0.97
mag and show no correlation with \mb, evidence of very different
behavior in the mid-UV which does not follow the one-parameter family
that dominates the optical luminosity-width relation.

\begin{figure} 
\resizebox{14cm}{!}{\includegraphics*{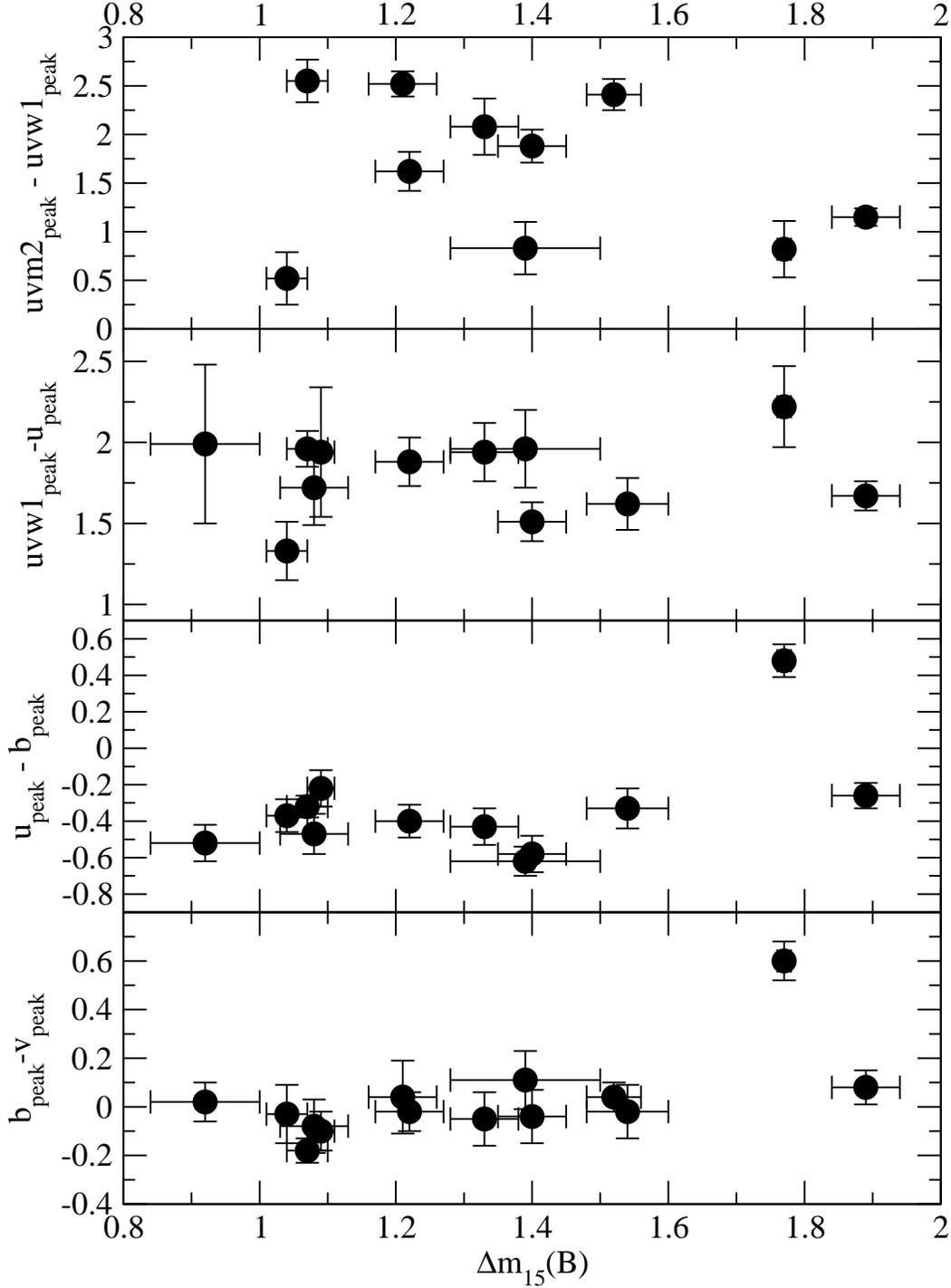}    }
\caption[UV colors plotted with respect to the optical decay rate \mb.
]  {Various extinction corrected colors plotted with respect to the
  optical decay rate \mb.  While many subluminous SNe~Ia can be
  identified in the optical by their very red colors, SN~2007on does
  not follow this trend, and the UV colors of both SNe 2005ke (which
  is very red in $b-v$ and $u-b$) and 2007on are similar to those of
  normal SNe~Ia.  The same behavior is seen when the host reddening is
  corrected with the CSLMC law.  }\label{plot_UVcolorsvdmb15}
\end{figure}

Considering colors of neighboring bands gives one a crude view of the
spectral shape and variability between objects.  These colors are
displayed in Figure \ref{plot_UVcolorsvdmb15}.  We see that
uvw1$_{rc}-u$ has a modestly larger scatter as does $u-b$, but that of
uvm2$-$uvw1$_{rc}$ is much larger.  While the differences are very
small in the 4000--6000~\AA\ ($bv$) regions, they become modestly
larger over the intervals 3000--5000~\AA\ and 2500--4000~\AA.  The
region shortward of $\sim2500$~\AA\ is found to be much more diverse.
Also of note is that while SN~2005ke is red in $b-v$, $u-b$, and
uvw1$_{rc}-u$, it appears normal in uvm2$-$uvw1$_{rc}$, suggesting
that the strong spectral differences do not extend as far in the UV.
In a comparison of the near-UV spectra of high-redshift SNe~Ia, both
\citet{Ellis_etal_2008} and \citet{Foley_etal_2008} found an increase
in the spectral diversity shortward of $\sim3700$~\AA.  Computing
colors for those spectra in the UVOT filters would allow the best
comparison of the dispersion and the UV shape
\citep{Sullivan_etal_2009} at low and high redshifts.

\subsection{Absolute Magnitudes} 
 
\begin{figure} 
\resizebox{11cm}{!}{\includegraphics*{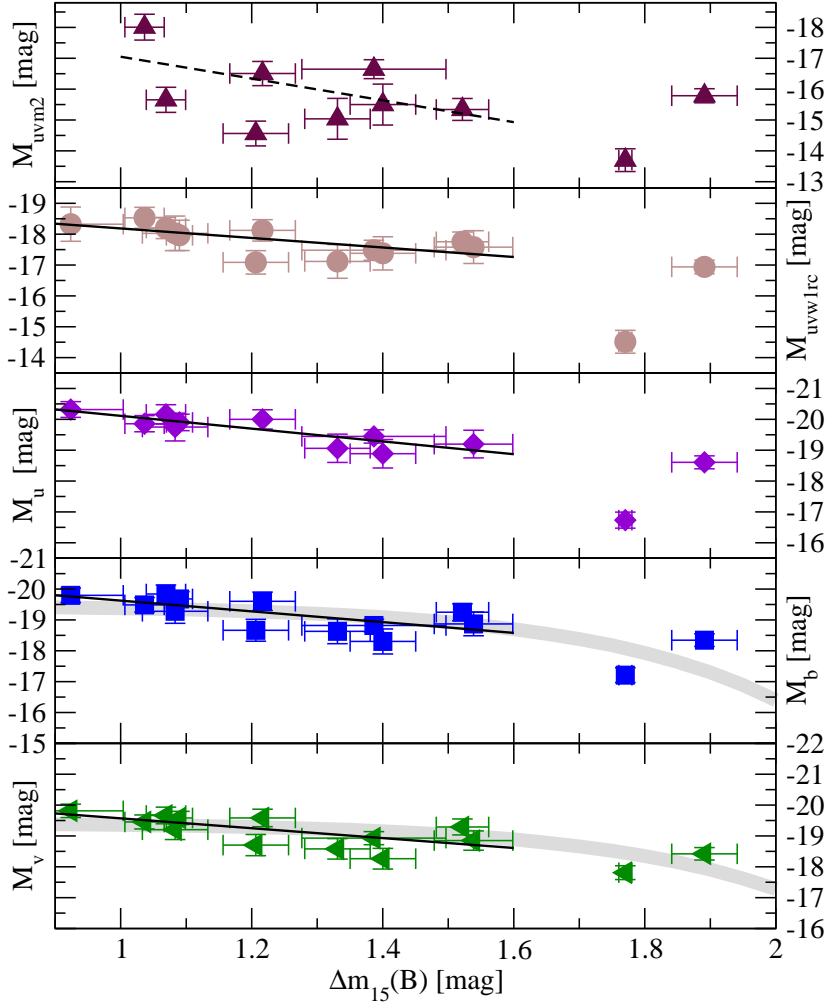}  }
\caption[UV and optical absolute magnitudes plotted with respect to
  their optical decay rate \mb.]  {UV and optical absolute magnitudes,
  calibrated with independent or Hubble-flow distances, plotted with
  respect to their optical decay rate \mb.  The host-galaxy reddening
  was corrected using the MW law.  Similar to the optical, the
  absolute magnitudes in the $u$ and uvw1 bands become slightly
  fainter with \mb\ for the SNe with $0.9 \leq \Delta m_{15}(B) \leq
  1.6$ mag.  Black lines show a fit to the absolute magnitudes with
  respect to \mb.  The absolute magnitudes in the uvm2 filter show
  considerably more scatter and are only poorly fit with \mb.
  SN~2005ke is even fainter in the optical and near-UV filters than
  one would expect based on its \mb, but SN~2007on is brighter.  The
  grey lines represent the \citet{Garnavich_etal_2004} absolute
  magnitudes in $B$ and $V$, with the dispersion of 0.2 mag
  represented by the width of the line.   }
\label{plot_absmagsmw}
\end{figure}

\begin{figure} 
\resizebox{11cm}{!}{\includegraphics*{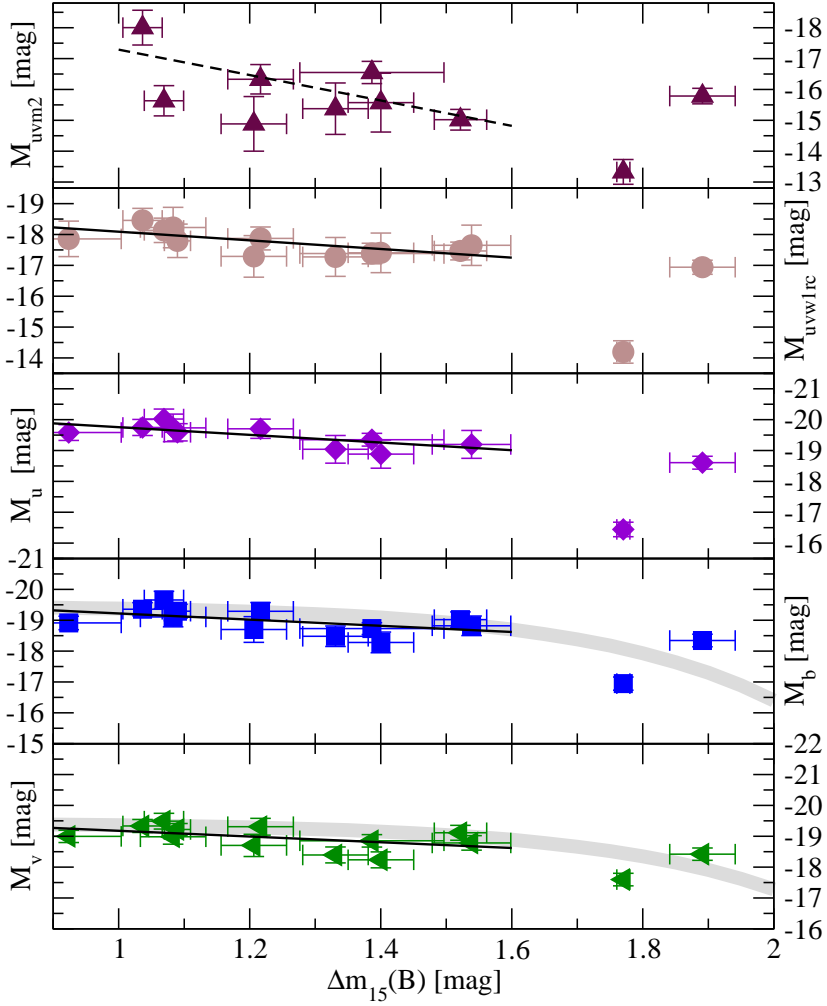}  }
\caption[UV and optical absolute magnitudes plotted with respect to
  their optical decay rate \mb.]  {UV and optical absolute magnitudes,
  calibrated with independent or Hubble-flow distances, plotted with
  respect to their optical decay rate \mb.  The host-galaxy reddening
  was corrected using the CSLMC law.  The plot is very similar to
  Figure \ref{plot_absmagsmw}, with the moderately extinguished
  SN~2007S on the left side being the only one that moves
  significantly compared to the others. 
}\label{plot_absmagscslmc}
\end{figure}

Tables \ref{table_absmagsmw} and \ref{table_absmags17} list the
absolute magnitudes for all of our SNe computed using the apparent
magnitudes of Table \ref{table_appmags}, the red-tail correction for
uvw1 of Table \ref{table_redtail}, the K-corrections of
Table~\ref{table_kcor}, the MW and host-galaxy reddenings (corrected
with the MW and CSLMC extinction laws) of Table~\ref{table_EBV}, and
the independent or Hubble-flow distance moduli of
Table~\ref{table_host}.  These absolute magnitudes are plotted in
Figures \ref{plot_absmagsmw} and \ref{plot_absmagscslmc} with respect
to the extinction-corrected optical decay rate \mb\ from either UVOT
or ground-based observations (also listed in Tables
\ref{table_absmagsmw} and \ref{table_absmags17} for easier
identification of the SNe).


The absolute magnitudes in the optical and UV, calculated with either
the MW or CSLMC extinction law, show similar behavior for low values
of \mb.  Within the range $0.9 < \Delta m_{15}(B) < 1.6$ mag, the $v$,
$b$, $u$, and uvw1$_{rc}$ absolute magnitudes scatter about the
average by $\sim 0.4$--0.5 mag (Table \ref{table_absmagsfits}).  The
uvm2 absolute magnitudes have a larger dispersion of $\sim 1$ mag.
Fitting the absolute magnitudes of the SNe with respect to \mb\ within
the range $0.9 < \Delta m_{15}(B) < 1.6$ mag, we find a slightly
steeper slope at shorter wavelengths, though the scatter in uvm2 is
especially large and the fit not well constrained.  The scatter about
the linear fit decreases to $\sim 0.3$--0.4 mag in the $v$, $b$, $u$,
and uvw1$_{rc}$ filters.  To estimate the intrinsic scatter along such
a linear relation, we tested how much scatter needed to be added in
quadrature to the estimated uncertainties in order for the reduced
$\chi^2$ to be near unity.  No intrinsic scatter is required for the
$v$, $b$, $u$, and uvw1$_{rc}$ filters --- the reduced $\chi^2$ values
are much less than unity, indicating that the observed scatter is
consistent with the estimated uncertainties and that those
uncertainties might be overestimated.  So while the $b$ and $v$
scatter apparent in Figures \ref{plot_absmagsmw} and
\ref{plot_absmagscslmc} is larger than that found for most other
samples of SNe~Ia (typically $\sim 0.2$ mag;
\citealp{Hamuy_etal_1996abs,Phillips_etal_1999,Altavilla_etal_2004,Garnavich_etal_2004}),
it is consistent with our estimated uncertainties.

In the uvm2, however, the absolute magnitudes do require an additional
scatter to have an acceptable reduced $\chi^2$.  The amount of
intrinsic dispersion required is 0.8 mag for the MW extinction law and
0.4 mag for the CSLMC law.  This reduction is likely due to the larger
uncertainties in the extinction due to the much larger value of
$R_{\rm uvm2}$.  That intrinsic scatter is required above the
observational errors is significant, considering that the optical
absolute magnitudes have intrinsic scatter masked by our errors.

Since a significant fraction of the scatter in the optical and near-UV
could come from the Hubble-flow distances, the absolute magnitudes
were also computed using the SN-derived MLCS31 (MLCS17) distances.  For
the subsample of normal SNe with MLCS31 (MLCS17) distances, this drops
the scatter in the absolute magnitudes in $v$, $b$, $u$, and
uvw1$_{rc}$ from 0.18, 0.23, 0.16, and 0.21 (0.20, 0.28, 0.19, and 0.25)
mag to 0.15, 0.15, 0.22, and 0.22 (0.11, 0.13, 0.22, and 0.22) mag,
respectively.  The scatter drops in the optical, as is expected for a
distance calculation based on the optical magnitudes, but the near-UV
scatter is not changed significantly.  The uvm2 scatter actually
increases slightly from 0.88 to 1.00 (0.90 to 0.99) mag, reinforcing
the conclusion that most of the scatter in uvm2 is intrinsic.
 
Our two SNe having high values of \mb\ are much harder to interpret.
SNe~2005ke is significantly fainter than the rest of the sample in the
UV filters as well as in the optical, where it is fainter than other
subluminous SNe with the same decay rate, represented by the
\citet{Garnavich_etal_2004} relation plotted in grey in Figures
\ref{plot_absmagsmw} and \ref{plot_absmagscslmc}. Conversely, SN~2007on is brighter than the
\citet{Garnavich_etal_2004} relation and in the UV has an absolute
magnitude comparable to that of the normal SNe~Ia.  Phillips (2009,
private communication) reports that spectroscopically SN~2007on
appears more similar to SNe having \mb\ $\approx 1.6$ mag, suggesting
that \mb\ is not a good discriminator for the most rapidly declining
SNe.  Recent work by \citet{Krisciunas_etal_2009} shows that SNe~Ia
may exhibit a bimodal distribution in the near-IR and optical absolute
magnitudes in this range of \mb, hence similar differences in the UV
should not be surprising.


\subsection{Search for Photometric UV Luminosity Indicators} 

The optical light of the highest-redshift SNe will be hard to observe,
so it would be valuable to have a distance-independent luminosity
indicator observable in the UV data to calibrate the magnitudes either
as a proxy for \mb\ or as a separate luminosity indicator.
Unfortunately, no such parameter has yet been found in our photometry.

\begin{figure} 
\resizebox{14cm}{!}{\includegraphics*{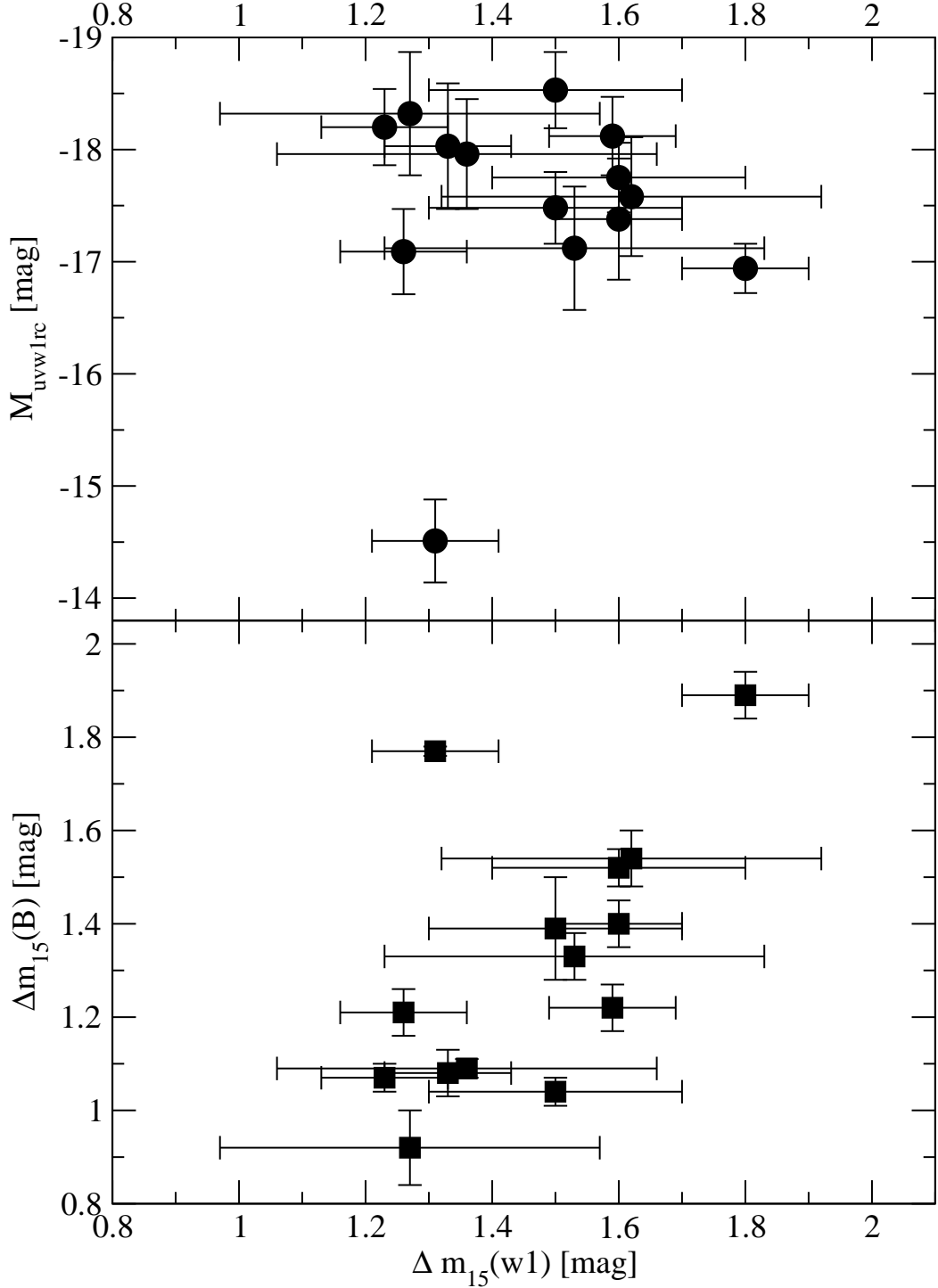}  } 
\caption[The uvw1 absolute magnitude and the optical decay rate
  \mb\ are plotted with respect to the uvw1 decay rate.]  {The uvw1
  absolute magnitude and the optical decay rate \mb\ are plotted with
  respect to the uvw1 decay rate.  The uvw1 absolute magnitudes have
  been corrected using the MW extinction law, but the same behavior is
  seen with the CSLMC law.  While the uvw1$_{rc}$ absolute magnitudes
  and \mwone\ both correlate weakly with \mb, there is no strong
  correlation between the absolute magnitudes and \mwone, which
  occupies a narrower range than its optical counterpart.
}\label{plot_uvw1decay}
\end{figure}

M10 parameterized the shapes of UV light curves in a manner similar to
that of \citet{Phillips_1993}, measuring the amount a supernova fades
in the first 15 days after maximum light.  However, as was noticed
after the first five SNe~Ia observed with {\it Swift}
\citep{Immler_etal_2006}, the early uvw1 light curves of all our
SNe~Ia are similar.  As shown in Figure \ref{plot_uvw1decay}, the
range of $\Delta m_{15}({\rm uvw1})$ is only $\sim 0.5$~mag (compared
to the $\sim 1$~mag difference between slow and fast decliners in the
optical).  Moreover, within that small range, there is no correlation
between $\Delta m_{15}({\rm uvw1})$ and $\Delta m_{15}(B)$ or the
uvw1$_{rc}$ absolute magnitude.  The range of decay rates is larger in
the uvm2 band, but again, the scatter in the absolute magnitude
vs. optical decay-rate relation is much stronger than any systematic
trend.  M10 do find a correlation between the optical decay rate and
the time that the light curve breaks from its steep to shallow decay,
but there is no clear correlation between the time of the break and
the UV absolute magnitudes.
 
\begin{figure} 
\resizebox{14cm}{!}{\includegraphics*{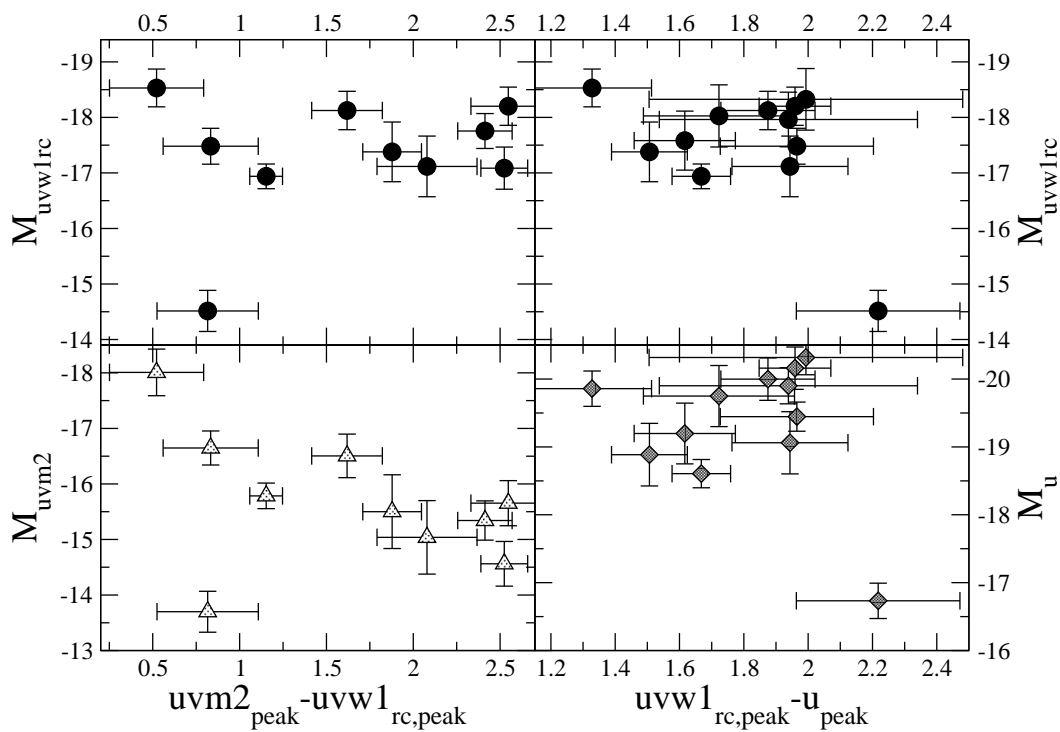}  } 
\caption[ ] {uvm2, uvw1$_{rc}$, and $u$ absolute magnitudes are
  plotted vs. uvm2$-$uvw1$_{rc}$ and uvw1$_{rc}-u$ colors.  There is
  no clear correlation except in uvm2, where the absolute magnitudes
  (with the exception of SN~2005ke) correlate with the colors because
  of the large dispersion in uvm2 compared to uvw1$_{rc}$.  The uvw1
  absolute magnitudes have been corrected using the MW extinction law,
  but the same behavior is seen with the CSLMC law.
}\label{plot_UVabsvUVcolors}
\end{figure}

Correlations between the absolute magnitudes and peak colors would be
most useful, since following a UV light curve as it fades would be
increasingly more difficult at higher redshifts.  The UV colors in our
small sample, however, do not show a strong correlation between the UV
colors and \mb, displayed in Figure \ref{plot_colorsvdmb15}.  Figure
\ref{plot_UVabsvUVcolors} directly compares the UV absolute magnitudes
vs.  UV colors, which is what would be very useful for measuring
distances using only UV observations.  There is no clear correlation
except in uvm2, where we essentially see the absolute magnitudes
correlate with the colors because of the large dispersion in uvm2
compared to uvw1$_{rc}$.

\begin{figure} 
\resizebox{12cm}{!}{\includegraphics*{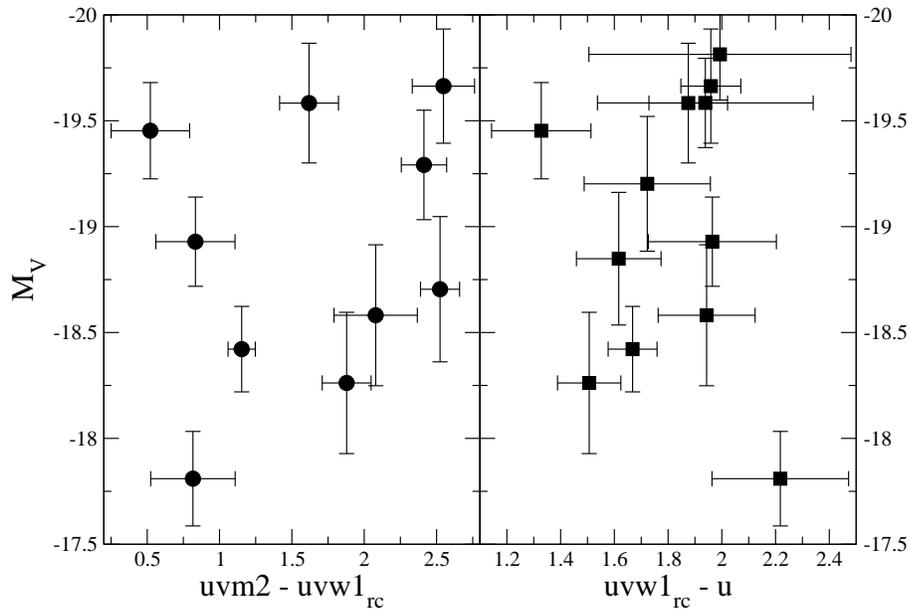}  } 
\caption[Absolute $V$ magnitudes plotted with respect to the UV colors.]  
  {Absolute $V$ magnitudes plotted with respect to the UV colors. No
  correlation is seen to correspond with the near-UV spectral
  correlation found by \citet{Foley_etal_2008UV}.  The uvw1 absolute
  magnitudes have been corrected using the MW extinction law, but the
  same behavior is seen with the CSLMC law.  }\label{plot_UVcolorsvV}
\end{figure}

\citet{Foley_etal_2008UV} have reported a possible correlation between
the peak luminosity of a SN~Ia and its spectral slope between
2770~\AA\ and 2900~\AA. While most {\sl Swift} grism spectra do not
have sufficiently high signal-to-noise ratio to measure this ratio
\citep{Bufano_etal_2009}, \citet{Foley_etal_2008UV} suggested that
UVOT colors might exhibit the same trend, but Figure
\ref{plot_UVcolorsvV} shows that the correlation is not seen in the
photometric colors of the larger bandpasses.  However, as the scatter
in our optical absolute magnitudes is consistent with the
observational errors, it would be difficult to find a correlation with the
intrinsic differences.  The broad-band filters may also wash out the
spectral features or exaggerate small errors in the extinction.


\section{Discussion} 

The UV absolute magnitudes of this sample of SNe~Ia seem to show two
distinct types of behavior.  At near-UV wavelengths (2500--4000~\AA)
probed by UVOT uvw1$_{rc}$ and $u$ and by ground-based $U$
\citep{Jha_etal_2006U}, the absolute magnitudes show a similar
relationship to the optical decay rate \mb\ as do the optical absolute
magnitudes.  As shown in Table \ref{table_absmagsfits}, for this
sample the scatter in the absolute magnitudes of SNe~Ia at peak is
comparable in the near-UV and optical.  However, this scatter is
higher than that found in the optical for larger samples (typically
$\sim 0.2$ mag;
\citealp{Hamuy_etal_1996abs,Phillips_etal_1999,Altavilla_etal_2004,Garnavich_etal_2004}),
and the UV scatter could be significantly higher.  If larger samples
put tighter limits on the intrinsic dispersion, the near-UV could be
useful for measuring luminosity distances.  This would be most
valuable at higher redshifts where fewer optical bands are available,
to put multi-color constraints on the reddening and distance.  How
useful they could be depends on the intrinsic scatter, which is hard
to measure in this sample due to its small size and large
observational uncertainties. While larger samples are now available in
$U$ \citep{Jha_etal_2006U,Hicken_etal_2009a}, the space-based UV
sample is still relatively small --- comparable to the 9 objects used
by \citet{Phillips_1993} to discover the peak absolute magnitude
vs. decline-rate relation.  Just as the systematics of optical
absolute magnitudes have been refined with larger samples of objects,
an increased number of UV measurements is needed to better understand
the behavior of SNe in the UV.  The origin and wavelength dependence
of extinction, and the uncertainty in the red-tail and K-corrections
due to intrinsic differences in the UV spectra, are needed to improve
the utility of the near-UV SN photometry.

Conversely, the absolute magnitudes in the mid-UV (2000--2500~\AA, the
UVOT uvm2 band) exhibit a large scatter and therefore are less useful
for distance determinations.  Studying the cause of that scatter in
nearby events, however, should assist us in understanding the SN~Ia
phenomenon, as well as the effect that different reddening laws or
host environments have on distance estimates.

Metallicity differences have been suggested as a possible second
parameter in the optical maximum magnitude vs. rate of decline
relation \citep[e.g.,][]{Mazzali_Podsiadlowski_2006}.  A
metallicity-dependent component to the absolute magnitudes could
result in an evolutionary bias as one observes higher-redshift SNe
with likely different metallicities\citep{Podsiadlowski_etal_2006}.  
A concern already for the
optical, this could be even more critical in the UV where the
metallicity plays a strong role in the reverse-fluorescence emission
\citep{Mazzali_2000,Sauer_etal_2008} as well as in the line blanketing
(see \citealp{Lentz_etal_2000, Dom_etal_2001}).  Moreover, a change in
metallicity could also bias SN-based extinction measurements which are
calibrated using the $B-V$ color of local objects
\citep{Dom_etal_2001}.  A better understanding could be reached
through detailed studies of the ages and metallicities of SN host
galaxies \citep{Gallagher_etal_2008} observed in the UV, and by a
search for correlations of such measures with the UV luminosity and
colors \citep{Sauer_etal_2008}.  If a correlation can be found, then
the UV data might provide leverage in probing the metallicity
dependence on the optical brightness and its evolution with redshift,
thus making SNe~Ia even better standardized candles in the optical.
There is disagreement as to whether metallicities have already been
found to affect optical measurements
\citep{Gallagher_etal_2008,Howell_etal_2009}.

In general, the measurements of SNe~Ia in the UV are hindered by the
low UV flux caused by metal line blanketing and extinction, yet the
strong sensitivity to these effects makes the UV region important for
constraining these parameters.  Thus, while rest-frame UV measurements
may not be ideal for cosmological distance measurements, the
understanding gleaned from UV data of local events will shed light on
the important effects of extinction and metallicity and how they might
evolve with redshift, thereby improving the reliability of SNe~Ia as
cosmological standard candles.


\acknowledgements

We are grateful to A. Goobar for extending his circumstellar
extinction model into the UV for us. This work is supported at Penn
State University by NASA contract NAS5-00136 and {\it Swift} Guest
Investigator grant NNH06ZDA001N. A.V.F. is grateful for the support of
NSF grant AST-0908886 and {\it Swift} Guest Investigator grant
NNX09AG54G.  This work made use of public data in the {\it Swift} data
archive and the NASA/IPAC Extragalactic Database (NED), which is
operated by the Jet Propulsion Laboratory, California Institute of
Technology, under contract with NASA.  The CfA Supernova Program at
Harvard University is supported by NSF grant AST-0907903.  The work of
M.T. is supported by grant number 2006022731 of the PRIN of Italian
Ministry of University and Science Research.


\appendix

\section{Red-Tail Corrections for Other Sources}

Vega magnitudes in the nominal UVOT filters and our red-tail-corrected
uvw2$_{rc}$ and uvw1$_{rc}$ were generated for a large sample of
different sources.  For stars we used the empirical stellar spectra
database of \citet{Pickles}.  Most of the Pickles spectra of later type stars 
(G and beyond) have zero flux in the UV, and those are excluded here.  
A representative UV-optical galaxy
spectrum for the main Hubble types from \citet{McCall_2004} was used,
together with synthetic spectra of a single burst of star formation 
viewed at one and fifteen Gyr computed using the model code P\'{E}GASE
\citep{PEGASE}.
Blackbody curves corresponding to temperatures ranging from 2,000 to
35,000~K were also used, as were {\it HST} UV spectra from a SN of
each major type: SNe 1992A (Ia; \citealp{Kirshner_etal_1993}), 1994I
(Ic; \citealp{Jeffery_etal_1994}), and 1999em (IIP;
\citealp{Baron_etal_2000}).  Various UV and UV-optical colors as well
as the red-leak correction factors were determined for each spectrum,
and these are tabulated for a subset of spectra in Table
\ref{table_redtailmore}.

\begin{figure} 
\resizebox{16cm}{!}{\includegraphics*{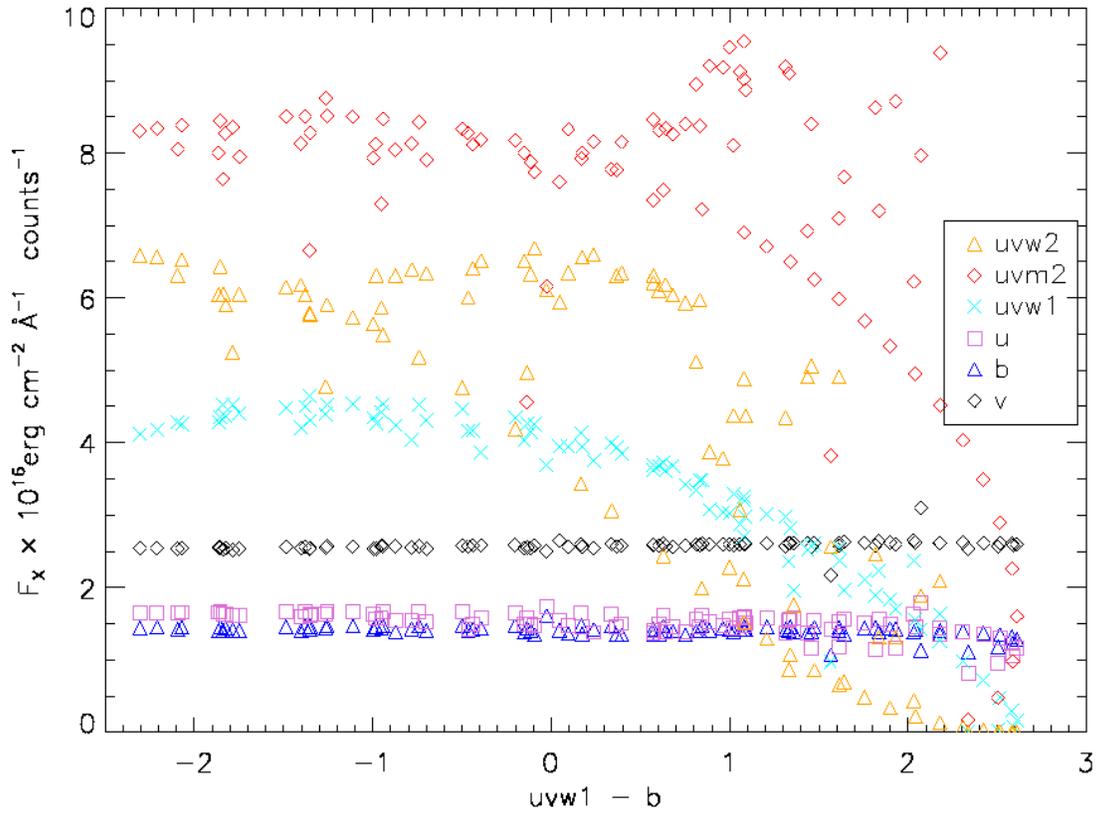} }
\caption[Flux conversion factors for the UVOT filters calculated for a
  variety of spectra.]  {Flux conversion factors for the UVOT filters
  calculated for a variety of spectra.  Multiplying the observed count rate 
in a given filter by the appropriate conversion factors gives the flux density 
the template spectrum has at the filter's effective wavelength for Vega.  The factors are fairly
  constant across different spectral types in the optical for this range of models but vary greatly in
  the UV.  } \label{plot_fluxfactors}
\end{figure}

The shape of the filter curves also strongly affects the calculation
of the flux density for different spectral shapes.  For comparison of
the flux of different objects, we keep the reference wavelength fixed
at the effective wavelengths determined for Vega
\citep{Poole_etal_2008} rather than allowing them to vary with the
spectral shape.  For each spectrum and filter, we determine the flux
conversion factor by dividing the average flux density in a
50~\AA\ window centered on the reference wavelength by the count rate
of the spectrum through that filter.  Thus, one multiplies the
observed count rate by the conversion factor to get the approximate
flux-density value.  \citet{Poole_etal_2008} presented average flux
conversion factors for a set of stellar and gamma-ray burst
(power-law) models.  Here we expand on this by giving the flux
conversion factors individually for the set of spectra in Table
\ref{table_fluxfactors}.  In Figure \ref{plot_fluxfactors} we show the
flux conversion factors varying strongly with the UV-optical color.
Accounting for this variation is critical for computing the UV flux
for red objects, as the average conversion factors assume an average
flux distribution, while for red objects the observed counts are
heavily biased to the red end of the filter curves.



\section{Extinction Coefficients}

\begin{figure} 
\resizebox{10cm}{!}{\includegraphics*{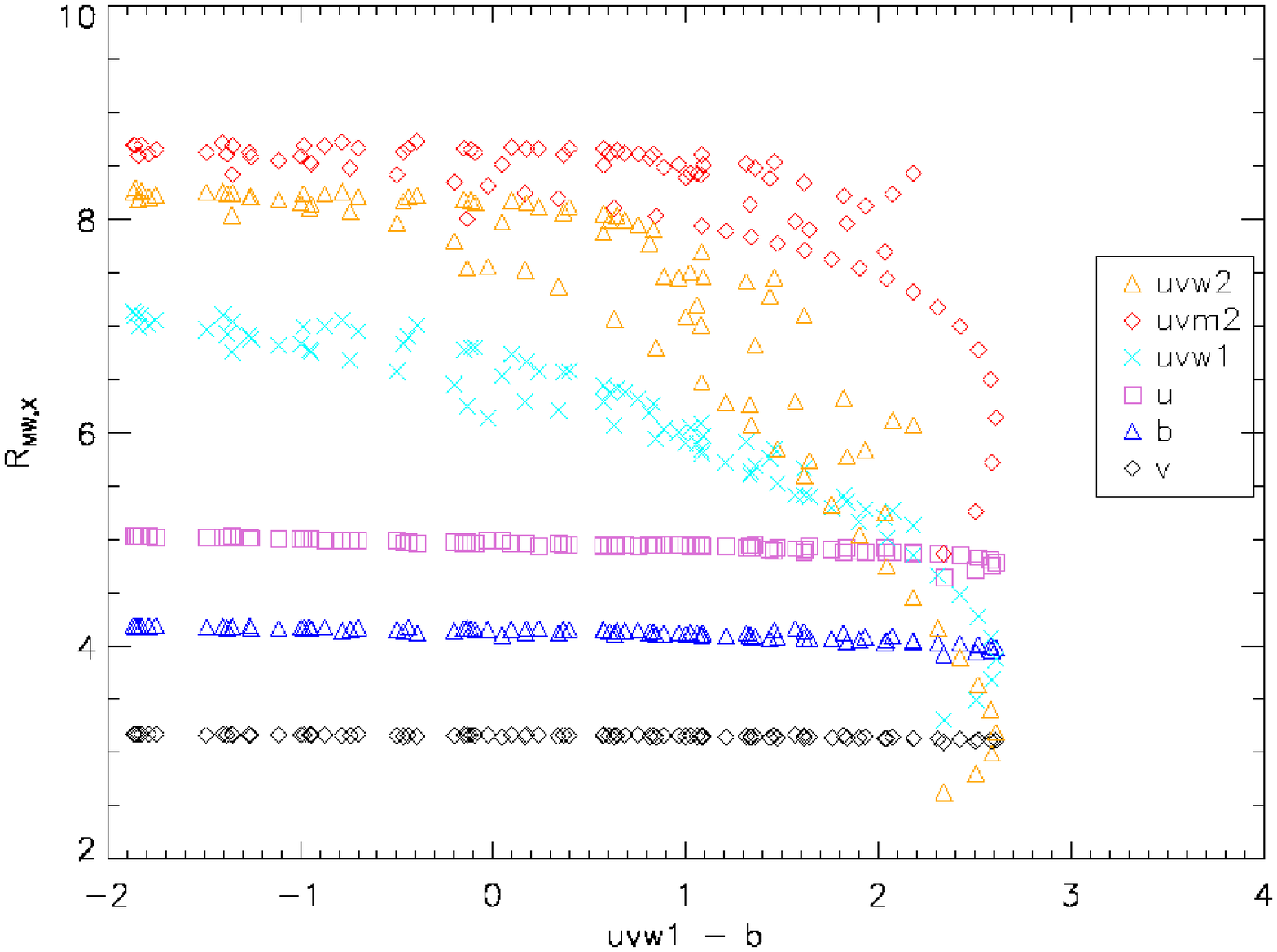}  } 
\caption[Extinction coefficients for the UVOT filters calculated for a
  variety of spectra.]  {Extinction coefficients for the UVOT filters
  calculated for a variety of spectra using the
  \citet{Cardelli_etal_1989} MW extinction law and $E(B-V) = 0.1$ mag.
  The uvw1$-b$ color on the abscissa is the unreddened color.  The factors are fairly
  constant in the optical for this range of models but vary greatly in
  the UV.  } \label{plot_rmwvalues}
\end{figure}

\begin{figure} 
\resizebox{10cm}{!}{\includegraphics*{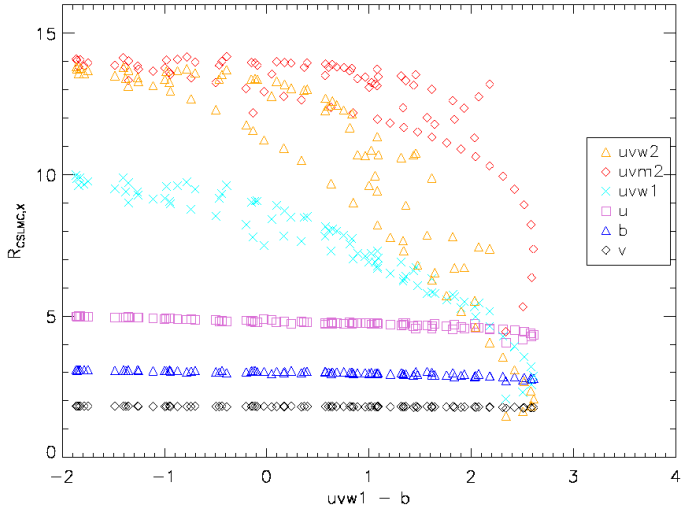}   }
\caption[Extinction coefficients for the UVOT filters calculated for a
  variety of spectra.]  {Extinction coefficients for the UVOT filters
  calculated for a variety of spectra using the
  \citet{Goobar_2008} circumstellar LMC extinction law and $E(B-V) = 0.1$ mag.
  The uvw1$-b$ color on the abscissa is the unreddened color.  The factors are fairly
  constant in the optical for this range of models but vary greatly in
  the UV.   } \label{plot_rcslmcvalues}
\end{figure}

Extinction coefficients are used to convert a measure of reddening to
the total extinction in a given filter X as in $A_X = R_X \times
E(B-V)$.  Since the extinction is a function of wavelength and the
filters have nonzero width, the effect of the reddening depends on the
spectral shape being extinguished.  Since in general extinction
increases to shorter wavelengths, a blue source is more efficiently
extinguished than a red source.  Below we extinguish the same set of
spectra by the \citep{Cardelli_etal_1989} $R_V = 3.1$ MW extinction
curve or the \citep{Goobar_2008} circumstellar LMC extinction law 
for $E(B-V) = 0.1$ mag.  The magnitude difference between the
extinguished and unextinguished spectra gives the extinction in each
filter.  The coefficients are plotted in Figures \ref{plot_rmwvalues} and \ref{plot_rcslmcvalues} and
tabulated in Tables \ref{table_rmwvalues} and \ref{table_rcslmcvalues}.



\begin{deluxetable}{lcccccccc} 
\rotate
\tablecaption{UVOT Filter Characteristics\label{table_uvotfilters}} 
\tablehead{ \colhead{Filter} & \colhead{$\lambda_{\rm central}$\tablenotemark{a}} & \colhead{FWHM} &  
\colhead{$\lambda_{\rm eff,Vega}$\tablenotemark{b}} & \colhead{$\lambda_{\rm eff,SNIa}$\tablenotemark{c}} & \colhead{$R_{\rm MW},1,X$} & \colhead{$R_{\rm MW},2,X$}& \colhead{  {\bf $R_{\rm CSLMC},1,X$} } & \colhead{  {\bf $R_{\rm CSLMC},2,X$}  }\\  
\colhead{} & \colhead{(\AA)} & \colhead{(\AA)}  & \colhead{(\AA)} & \colhead{(\AA)}  & \colhead{(mag)} & \colhead{(mag)}   & \colhead{(mag)} & \colhead{(mag)}  
}  
\tablewidth{0pt} 
\startdata 
uvw2        & 1941 & 556 & 2030 & 3064 & 6.20 & -0.95& 6.77 & -2.07\\ 
uvw2$_{rc}$ & 1941 & 556 & \nodata & 2010 & 8.60 & -0.18& 14.58 & -0.75\\ 
uvm2        & 2248 & 514 & 2231 & 2360 & 8.01 & -0.85& 11.33 & -3.13\\ 
uvw1        & 2605 & 653 & 2634 & 3050 & 5.43 & -0.41& 5.58 & -1.17\\ 
uvw1$_{rc}$ & 2605 & 653 & \nodata & 2890 & 6.06 & -0.13& 7.33 & -0.39\\ 
$u$         & 3464 & 787 & 3501 & 3600 & 4.92 & -0.03& 4.68 & -0.17\\ 
$b$         & 4371 & 982 & 4329 & 4340 & 4.16 & -0.04& 3.06 & -0.09\\ 
$v$         & 5441 & 730 & 5402 & 5400 & 3.16 & -0.01& 1.81 & -0.01\\ 
\enddata 
\tablenotetext{a}{$\lambda_{\rm central}$ is the wavelength midway between the 
wavelengths at which the effective area is equal to half the maximum effective area.}
\tablenotetext{b}{$\lambda_{\rm eff,Vega}$ refers to the photon-weighted effective  
wavelength of the Vega spectrum after passing through the filter from \citet{Poole_etal_2008}. }  
\tablenotetext{c}{$\lambda_{\rm eff,SNIa}$ refers to the photon-weighted effective  
wavelength of the SN~1992A after passing through the filter.   
$R_{1,X}$ and  $R_{2,X}$  are the extinction coefficients  
such that $A_X = R_{1,X} E(B-V) + R_{2,X} E(B-V)^2 $.  
 They were calculated using the SN~1992A spectrum at $\sim 5$ days
after maximum brightness, extinguished by the Milky Way (MW)
extinction law with $R_V = 3.1$ \citep{Cardelli_etal_1989} or the
circumstellar Large Magellanic Cloud (LMC) extinction law from
\citep{Goobar_2008}.
They should not be used to  
calculate the extinction to arbitrary sources observed with UVOT, 
and these terms can vary even for SNe~Ia.  } 
\end{deluxetable} 

\begin{deluxetable}{lllllllll} 
\tabletypesize{\small} 
\rotate 
\tablecaption{Apparent Magnitudes at Maximum Light \label{table_appmags}  } 
\tablehead{ 
\colhead{Name} & \colhead{\mb}  & \colhead{uvw2\tablenotemark{a}} & \colhead{uvm2} & \colhead{uvw1}  &  
\colhead{$u$}  & \colhead{$b$}  & \colhead{$v$}   & \colhead{ref\tablenotemark{b} } \\ 
\colhead{  } & \colhead{ (mag)   } & \colhead{ (mag)   } & \colhead{  (mag)  } & \colhead{  (mag)  } & \colhead{ (mag) } & 
\colhead{(mag) } & \colhead{(mag) } & \colhead{ }  
} 
\tablewidth{0pt} 
\startdata 

SN 2005am &   1.52 $\pm$   0.04 &  16.95 $\pm$   0.08 &  18.26 $\pm$   0.13 &  15.35 $\pm$  0.06 &   \nodata           &  13.90 $\pm$   0.04 &  13.75 $\pm$  0.03 & 1,2     \\ 
SN 2005cf &   1.07 $\pm$   0.03 &  16.83 $\pm$   0.08 &  18.32 $\pm$   0.19 &  15.11 $\pm$  0.07 &  13.41 $\pm$   0.05 &  13.54 $\pm$   0.02 &  13.53 $\pm$  0.02 & 3,3       \\ 
SN 2005df &   1.21 $\pm$   0.05 &  15.61 $\pm$   0.07 &  16.85 $\pm$   0.09 &  13.90 $\pm$  0.07 &   \nodata           &  12.50 $\pm$   0.10 &  12.40 $\pm$  0.10 & 4,4         \\
SN 2005ke &   1.77 $\pm$   0.01 &  18.41 $\pm$   0.11 &  18.87 $\pm$   0.15 &  17.09 $\pm$  0.09 &  15.52 $\pm$   0.07 &  14.92 $\pm$   0.05 &  14.22 $\pm$  0.05 & 5,U       \\ 
SN 2006dm &   1.54 $\pm$   0.06 &  18.93 $\pm$   0.13 &   \nodata          &  17.63 $\pm$  0.09 &  15.99 $\pm$   0.07 &  16.17 $\pm$   0.05 &  16.13 $\pm$  0.05 & 1,U        \\ 
SN 2006ej &   1.39 $\pm$   0.11 &  18.73 $\pm$   0.13 &  18.46 $\pm$   0.14 &  17.00 $\pm$  0.08 &  15.47 $\pm$   0.06 &  15.93 $\pm$   0.05 &  15.80 $\pm$  0.10 & 1,U        \\ 
SN 2007S  &   0.92 $\pm$   0.08 &   \nodata           &   \nodata         &  17.80 $\pm$  0.14 &  15.86 $\pm$   0.07 &  15.94 $\pm$   0.05 &  15.53 $\pm$  0.05 & U,U  \\ 
SN 2007af &   1.22 $\pm$   0.05 &  16.50 $\pm$   0.09 &  17.12 $\pm$   0.15 &  14.77 $\pm$  0.07 &  13.16 $\pm$   0.07 &  13.39 $\pm$   0.05 &  13.25 $\pm$  0.05 & U,U       \\ 
SN 2007co &   1.09 $\pm$   0.02 &  20.11 $\pm$   0.20 &   \nodata          &  18.80 $\pm$  0.15 &  16.99 $\pm$   0.07 &  16.86 $\pm$   0.05 &  16.69 $\pm$  0.05 & U,U       \\ 
SN 2007cq &   1.04 $\pm$   0.03 &  18.61 $\pm$   0.13 &  18.53 $\pm$   0.19 &  17.53 $\pm$  0.11 &   16.12 $\pm$  0.06  &  16.27 $\pm$   0.05 &  16.17 $\pm$  0.10 & 2,U        \\
SN 2007cv &   1.33 $\pm$   0.05 &  18.47 $\pm$   0.13 &  19.60 $\pm$   0.22 &  16.83 $\pm$  0.09 &  15.08 $\pm$   0.06 &  15.30 $\pm$   0.05 &  15.15 $\pm$  0.05 & U,U         \\
SN 2007on &   1.89 $\pm$   0.05 &  15.65 $\pm$   0.05 &  15.78 $\pm$   0.05 &  14.36 $\pm$  0.05 &  12.93 $\pm$   0.05 &  13.14 $\pm$   0.05 &  13.06 $\pm$  0.05 & 5,U       \\ 
SN 2008Q  &   1.40 $\pm$   0.05 &  16.42 $\pm$   0.05 &  17.09 $\pm$   0.08 &  14.84 $\pm$  0.05 &  13.39 $\pm$   0.05 &  13.85 $\pm$   0.05 &  13.80 $\pm$  0.05 & U,U         \\
SN 2008ec &   1.08 $\pm$   0.05 &  18.86 $\pm$   0.17 &   \nodata          &  17.30 $\pm$  0.11 &  15.62 $\pm$   0.07 &  15.83 $\pm$   0.05 &  15.70 $\pm$  0.05 & U,U            \\ 

\enddata 
\tablenotetext{a}{1$\sigma$ errors.} 

\tablenotetext{b}{References are given for the \mb\ and $bv$ peak
  magnitudes while the UV maxima are all UVOT measurements from
  \citep{Milne_etal_2010}.  }
 \tablerefs{(U) UVOT, M10; (1) Li et al. 2006; (2) KAIT, M10; (3)
  Pastorello et al. 2007; (4) ANU, M10; (5) CSP, Stritzinger et al.,
  in preparation.
} 
\end{deluxetable} 


\begin{deluxetable}{lcrcccc} 
\tablecaption{Red-Tail Corrections of SNe~Ia near Maximum Light \label{table_redtail}} 
\tablehead{ \colhead{SN} & \colhead{w2$_{rc}$ fraction\tablenotemark{a}} & \colhead{rc$_{w2}$\tablenotemark{b}} &  
 \colhead{w1$_{rc}$ fraction} & \colhead{rc$_{w1}$}  \\  
\colhead{} & \colhead{} & \colhead{(mag)} & \colhead{(mag)} & \colhead{} 
}  
\tablewidth{0pt} 
\startdata 

SN~2005am &            0.08 &       -2.35 $\pm$    0.31 &        0.59 &       -0.36 $\pm$    0.05 \\    
SN~2005cf &            0.04 &       -2.78 $\pm$    0.38 &        0.51 &       -0.47 $\pm$    0.06 \\    
SN~2005df &            0.10 &       -2.26 $\pm$    0.27 &        0.62 &       -0.35 $\pm$    0.06 \\    
SN~2005ke &            0.22 &       -1.28 $\pm$    0.23 &        0.39 &       -0.79 $\pm$    0.22 \\    
SN~2006dm &            0.20 &       -1.50 $\pm$    0.26 &        0.66 &       -0.26 $\pm$    0.07 \\    
SN~2006ej &            0.45 &       -0.63 $\pm$    0.17 &        0.44 &       -0.69 $\pm$    0.21 \\    
SN~2007S  &            0.11 &       -1.08 $\pm$    0.39 &        0.29 &       -0.82 $\pm$    0.46 \\    
SN~2007af &            0.12 &       -1.69 $\pm$    0.25 &        0.51 &       -0.47 $\pm$    0.10 \\    
SN~2007co &            0.16 &       -1.16 $\pm$    0.45 &        0.37 &       -0.71 $\pm$    0.36 \\    
SN~2007cq &            0.42 &       -0.56 $\pm$    0.17 &        0.60 &       -0.30 $\pm$    0.12 \\    
SN~2007cv &            0.07 &       -2.29 $\pm$    0.41 &        0.51 &       -0.45 $\pm$    0.12 \\    
SN~2007on &            0.43 &       -0.80 $\pm$    0.10 &        0.67 &       -0.28 $\pm$    0.05 \\    
SN~2008Q  &            0.20 &       -1.46 $\pm$    0.17 &        0.68 &       -0.22 $\pm$    0.04 \\    
SN~2008ec &            0.10 &       -1.81 $\pm$    0.31 &        0.53 &       -0.41 $\pm$    0.18 \\    

\enddata 
\tablenotetext{a}{w2$_{rc}$ and w1$_{rc}$ fractions refer to the fraction 
of the counts in the uvw2 (uvw1) filter 
that would pass through the red-tail corrected uvw2$_{rc}$ (uvw1$_{rc}$) filter.  }
\tablenotetext{b}{rc$_X$ is the magnitude difference, 
including a zeropoint correction, between synthetic photometry in the nominal 
filter and with the red-tail cutoff.  The $1\sigma$ error is estimated by offsetting the observed 
magnitudes by the $1\sigma$ errors and repeating the spectral mangling 
and synthetic spectrophotometry. } 
\end{deluxetable} 
\clearpage

\begin{figure} 
\plottwo{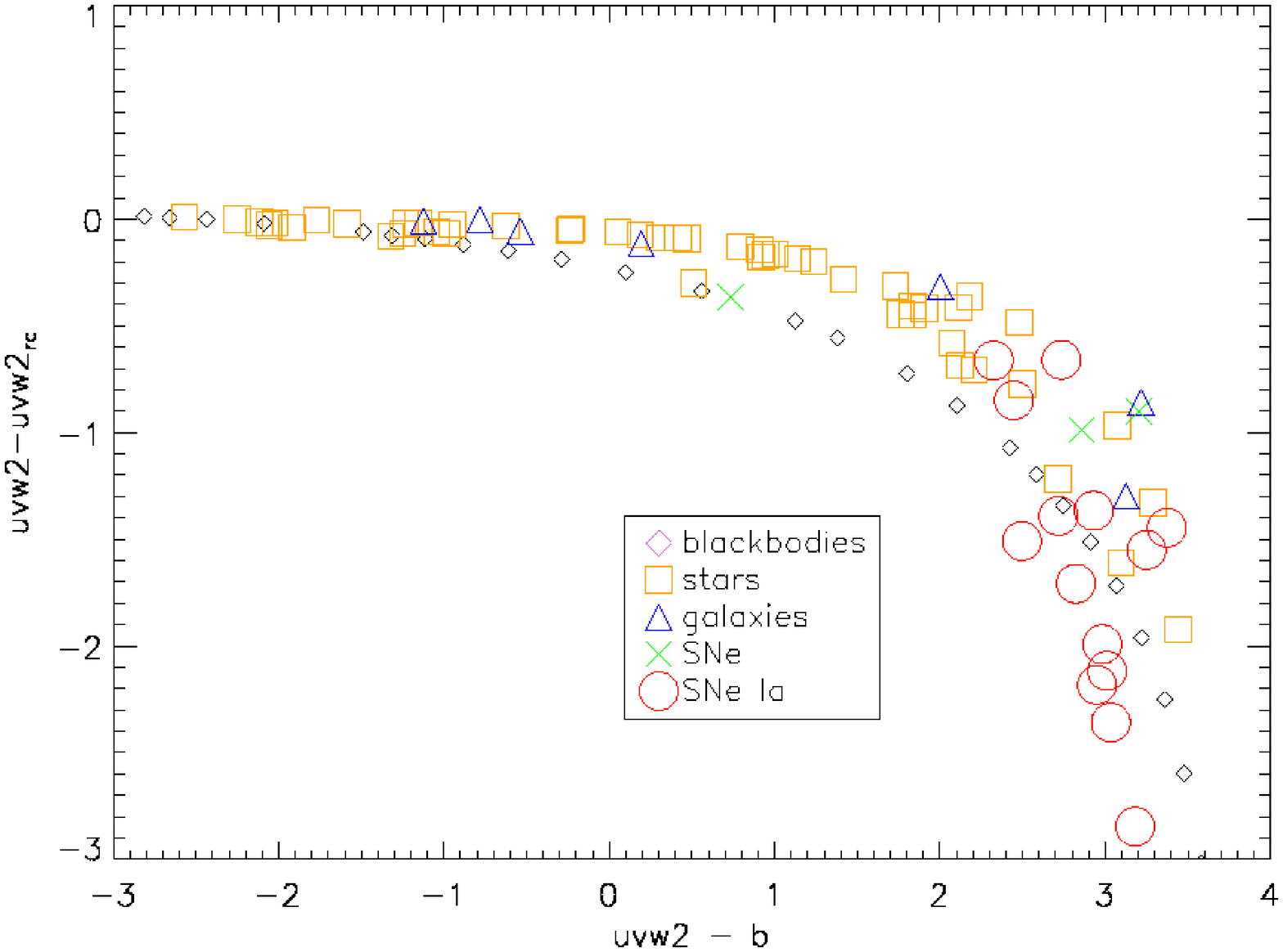}{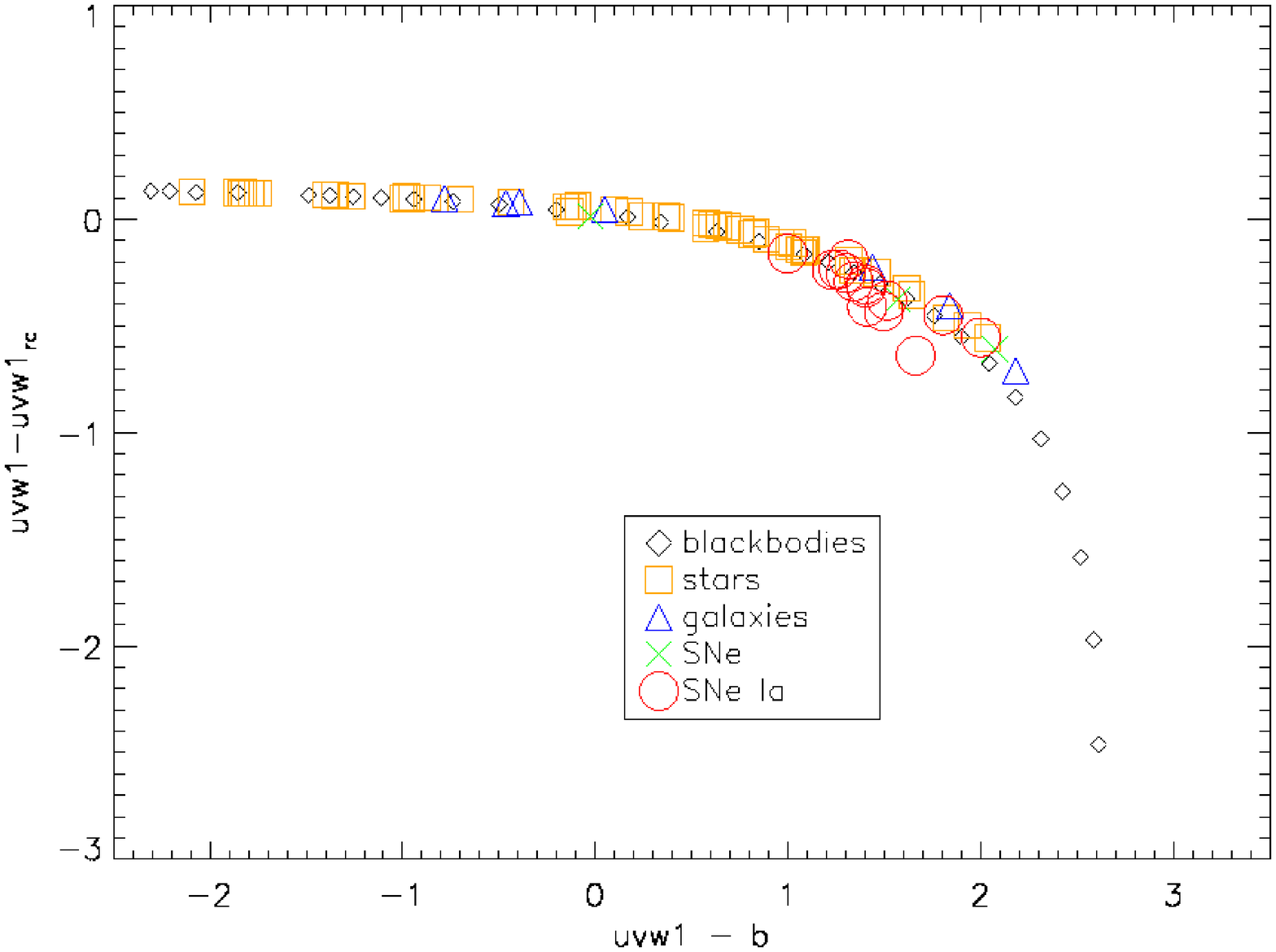}
\caption[Red-tail corrections for the uvw2 and uvw1 filters.]
        {Red-tail corrections for the uvw2 and uvw1 filters plotted
          with respect to the uvw2$-b$ and uvw1$-b$ colors for a variety
          of object classes.  } \label{plot_redcor}
\end{figure} 


\begin{deluxetable}{lrrrrr} 
\tabletypesize{\small} 
\tablecaption{K-Corrections for SNe~Ia Near Maximum Light \label{table_kcor}  } 
\tablehead{ 
\colhead{Name}  & \colhead{K$_{\rm uvm2}$}   &  \colhead{K$_{\rm w1rc}$}  & \colhead{K$_u$}  
& \colhead{K$_b$} & \colhead{K$_v$}  \\ 
 \colhead{ (mag)   } & \colhead{ (mag)   } & \colhead{  (mag)  } & \colhead{  (mag)  } &  
 \colhead{(mag) } & \colhead{ (mag)}   
} 
\tablewidth{0pt} 
\startdata 

SN 2005am &        0.12 $\pm$    0.01 &        0.10 $\pm$    0.01 &        0.04 $\pm$    0.01 &       -0.01 $\pm$    0.01 &       -0.00 $\pm$    0.01 \\    
SN 2005cf &        0.10 $\pm$    0.01 &        0.09 $\pm$    0.01 &        0.04 $\pm$    0.01 &       -0.01 $\pm$    0.01 &       -0.00 $\pm$    0.01 \\    
SN 2005df &        0.06 $\pm$    0.01 &        0.05 $\pm$    0.01 &        0.02 $\pm$    0.01 &       -0.01 $\pm$    0.01 &       -0.00 $\pm$    0.01 \\    
SN 2005ke &        0.02 $\pm$    0.01 &        0.05 $\pm$    0.01 &        0.05 $\pm$    0.01 &        0.00 $\pm$    0.01 &        0.00 $\pm$    0.01 \\    
SN 2006dm &        0.26 $\pm$    0.08 &        0.25 $\pm$    0.01 &        0.06 $\pm$    0.01 &       -0.02 $\pm$    0.01 &        0.00 $\pm$    0.01 \\    
SN 2006ej &        0.01 $\pm$    0.03 &        0.22 $\pm$    0.02 &        0.08 $\pm$    0.01 &       -0.04 $\pm$    0.01 &        0.01 $\pm$    0.01 \\    
SN 2007S  &        0.07 $\pm$    0.10 &        0.28 $\pm$    0.04 &        0.11 $\pm$    0.03 &       -0.01 $\pm$    0.01 &        0.04 $\pm$    0.01 \\    
SN 2007af &        0.04 $\pm$    0.01 &        0.07 $\pm$    0.01 &        0.03 $\pm$    0.01 &       -0.01 $\pm$    0.01 &       -0.00 $\pm$    0.01 \\    
SN 2007co &        0.12 $\pm$    0.13 &        0.31 $\pm$    0.03 &        0.12 $\pm$    0.03 &       -0.01 $\pm$    0.01 &        0.02 $\pm$    0.01 \\    
SN 2007cq &        0.05 $\pm$    0.06 &        0.21 $\pm$    0.02 &        0.07 $\pm$    0.02 &       -0.02 $\pm$    0.01 &        0.01 $\pm$    0.01 \\    
SN 2007cv &        0.10 $\pm$    0.02 &        0.10 $\pm$    0.01 &        0.04 $\pm$    0.01 &       -0.01 $\pm$    0.01 &       -0.00 $\pm$    0.01 \\    
SN 2007on &        0.02 $\pm$    0.01 &        0.06 $\pm$    0.01 &        0.03 $\pm$    0.01 &       -0.01 $\pm$    0.01 &       -0.00 $\pm$    0.01 \\    
SN 2008Q  &        0.08 $\pm$    0.01 &        0.09 $\pm$    0.01 &        0.03 $\pm$    0.01 &       -0.02 $\pm$    0.01 &       -0.00 $\pm$    0.01 \\    
SN 2008ec &        0.18 $\pm$    0.06 &        0.21 $\pm$    0.01 &        0.06 $\pm$    0.01 &       -0.02 $\pm$    0.01 &        0.00 $\pm$    0.01 \\    

 \enddata  
\end{deluxetable} 

\begin{deluxetable}{lrrrrrr} 
\tabletypesize{\small} 
\tablecaption{$E(B-V)$ Color-Excess Measurements\label{table_EBV} }  
\tablehead{ 
\colhead{Name} & \colhead{MW\tablenotemark{a,b}}  & \colhead{Peak\tablenotemark{c}}  & \colhead{Tail\tablenotemark{d}}  &  
\colhead{MLCS31\tablenotemark{e}}  &\colhead{MLCS17\tablenotemark{f}} 
} 
\tablewidth{0pt} 
\startdata 
SN 2005am & 0.05 $\pm$   0.01 &   0.13 $\pm$   0.06 &   0.02 $\pm$   0.05 &   0.06 $\pm$   0.03 &   0.01 $\pm$   0.02 \\
SN 2005cf & 0.10 $\pm$   0.02 &  -0.01 $\pm$   0.05 &   0.16 $\pm$   0.04 &   0.09 $\pm$   0.04 &   0.07 $\pm$   0.04 \\
SN 2005df & 0.03 $\pm$   0.01 &   0.13 $\pm$   0.15 &   0.02 $\pm$   0.10 &   0.03 $\pm$   0.03 &   \nodata          \\
SN 2005ke & 0.02 $\pm$   0.01 &   0.44 $\pm$   0.08 &   0.04 $\pm$   0.03 &   0.08 $\pm$   0.03 &   0.02 $\pm$   0.02 \\
SN 2006dm & 0.04 $\pm$   0.01 &   0.05 $\pm$   0.08 &  -0.03 $\pm$   0.14 &   \nodata          &   \nodata          \\
SN 2006ej & 0.04 $\pm$   0.01 &   0.18 $\pm$   0.12 &   0.32 $\pm$   0.19 &   0.03 $\pm$   0.02 &   0.01 $\pm$   0.02 \\
SN 2007S  & 0.03 $\pm$   0.01 &   0.53 $\pm$   0.08 &   0.38 $\pm$   0.05 &   0.41 $\pm$   0.03 &   0.27 $\pm$   0.03 \\
SN 2007af & 0.04 $\pm$   0.01 &   0.17 $\pm$   0.08 &   0.09 $\pm$   0.05 &   0.13 $\pm$   0.03 &   0.07 $\pm$   0.03 \\
SN 2007co & 0.11 $\pm$   0.02 &   0.16 $\pm$   0.08 &   \nodata          &   0.19 $\pm$   0.04 &   0.13 $\pm$   0.04 \\
SN 2007cq & 0.11 $\pm$   0.02 &   0.10 $\pm$   0.12 &   \nodata          &   0.06 $\pm$   0.04 &   0.04 $\pm$   0.03 \\
SN 2007cv & 0.07 $\pm$   0.01 &   0.14 $\pm$   0.08 &   \nodata          &   \nodata          &   \nodata          \\
SN 2007on & 0.01 $\pm$   0.01 &  -0.45 $\pm$   0.08 &  -0.11 $\pm$   0.08 &   \nodata          &   \nodata          \\
SN 2008Q  & 0.08 $\pm$   0.01 &   0.02 $\pm$   0.08 &   \nodata          &   \nodata          &   \nodata          \\
SN 2008ec & 0.07 $\pm$   0.01 &   0.16 $\pm$   0.08 &   \nodata          &   \nodata          &   \nodata          \\

 \enddata 
\tablenotetext{a}{Milky Way $E(B-V)$ value from \citet{Schlegel_etal_1998}. } 
\tablenotetext{b}{1$\sigma$ errors. } 
\tablenotetext{c}{Host $E(B-V)$ determined from the peak magnitudes
  above according to the intrinsic $B-V$ versus \mb\ relations of
  \citet{Phillips_etal_1999} and \citet{Garnavich_etal_2004}. }
\tablenotetext{d}{Host $E(B-V)$ determined using the Lira relation
  and the $B-V$ colors 30--60 days after maximum light
  \citep{Phillips_etal_1999}. }
\tablenotetext{e}{Host $E(B-V)$ derived from MLCS2k2 fitting
  \citep{Jha_etal_2007} using $R_V=3.1$ \citep{Hicken_etal_2009b}.  }
\tablenotetext{f}{Host $E(B-V)$ derived from MLCS2k2 fitting
  \citep{Jha_etal_2007} using $R_V=1.7$ \citep{Hicken_etal_2009b}.  }

\end{deluxetable} 

\begin{deluxetable}{llrrrrrr} 
\tabletypesize{\small} 
\tablecaption{Extinction Coefficients\label{table_rvalues} }  
\tablehead{ 
\colhead{Name} & \colhead{Reddening\tablenotemark{a}} & \colhead{$R_{\rm uvm2}$\tablenotemark{b} } & \colhead{$R_{\rm uvw1rc}$} & \colhead{$R_u$} & \colhead{$R_b$} & \colhead{$R_v$ }
} 
\tablewidth{0pt} 
\startdata 

SN~2005am & MW MW & 6.96 $\pm$    0.16 &        5.86 $\pm$    0.02 &        4.95 $\pm$    0.02 &        4.17 $\pm$    0.00 &        3.16 $\pm$    0.00 \\    
SN~2005cf & MW MW & 6.78 $\pm$    0.24 &        5.83 $\pm$    0.03 &        4.93 $\pm$    0.02 &        4.17 $\pm$    0.00 &        3.16 $\pm$    0.00 \\    
SN~2005df & MW MW & 7.07 $\pm$    0.15 &        5.89 $\pm$    0.02 &        4.95 $\pm$    0.02 &        4.17 $\pm$    0.01 &        3.16 $\pm$    0.01 \\    
SN~2005ke & MW MW & 8.28 $\pm$    0.31 &        6.20 $\pm$    0.18 &        4.90 $\pm$    0.02 &        4.12 $\pm$    0.01 &        3.15 $\pm$    0.00 \\    
SN~2006dm & MW MW & 7.34 $\pm$    0.40 &        5.95 $\pm$    0.06 &        4.96 $\pm$    0.02 &        4.16 $\pm$    0.01 &        3.16 $\pm$    0.00 \\    
SN~2006ej & MW MW & 8.84 $\pm$    0.23 &        6.57 $\pm$    0.35 &        4.91 $\pm$    0.02 &        4.18 $\pm$    0.01 &        3.15 $\pm$    0.00 \\    
SN~2007S  & MW MW & 8.28 $\pm$    0.53 &        6.40 $\pm$    0.62 &        4.91 $\pm$    0.04 &        4.14 $\pm$    0.01 &        3.15 $\pm$    0.00 \\    
SN~2007af & MW MW & 7.77 $\pm$    0.25 &        5.96 $\pm$    0.05 &        4.93 $\pm$    0.02 &        4.18 $\pm$    0.01 &        3.16 $\pm$    0.00 \\    
SN~2007co & MW MW & 8.13 $\pm$    0.61 &        6.25 $\pm$    0.50 &        4.91 $\pm$    0.04 &        4.13 $\pm$    0.01 &        3.15 $\pm$    0.00 \\    
SN~2007cq & MW MW & 8.46 $\pm$    0.29 &        6.37 $\pm$    0.20 &        4.94 $\pm$    0.03 &        4.15 $\pm$    0.01 &        3.15 $\pm$    0.00 \\    
SN~2007cv & MW MW & 7.13 $\pm$    0.32 &        5.86 $\pm$    0.04 &        4.94 $\pm$    0.02 &        4.17 $\pm$    0.01 &        3.16 $\pm$    0.00 \\    
SN~2007on & MW MW & 8.35 $\pm$    0.10 &        6.20 $\pm$    0.04 &        4.95 $\pm$    0.02 &        4.18 $\pm$    0.01 &        3.16 $\pm$    0.00 \\    
SN~2008Q  & MW MW & 7.56 $\pm$    0.13 &        5.96 $\pm$    0.02 &        4.96 $\pm$    0.02 &        4.19 $\pm$    0.01 &        3.16 $\pm$    0.00 \\    
SN~2008ec & MW MW & 7.40 $\pm$    0.44 &        5.91 $\pm$    0.06 &        4.94 $\pm$    0.03 &        4.16 $\pm$    0.01 &        3.15 $\pm$    0.00 \\    
SN~2005am & host MW & 6.94 $\pm$    0.15 &        5.90 $\pm$    0.02 &        4.97 $\pm$    0.02 &        4.21 $\pm$    0.00 &        3.19 $\pm$    0.00 \\    
SN~2005cf & host MW & 6.66 $\pm$    0.22 &        5.73 $\pm$    0.04 &        4.93 $\pm$    0.02 &        4.20 $\pm$    0.00 &        3.18 $\pm$    0.00 \\    
SN~2005df & host MW & 7.06 $\pm$    0.14 &        5.91 $\pm$    0.02 &        4.96 $\pm$    0.02 &        4.19 $\pm$    0.01 &        3.18 $\pm$    0.01 \\    
SN~2005ke & host MW & 8.23 $\pm$    0.32 &        6.19 $\pm$    0.16 &        4.91 $\pm$    0.02 &        4.13 $\pm$    0.01 &        3.17 $\pm$    0.00 \\    
SN~2006dm & host MW & 7.47 $\pm$    0.39 &        6.09 $\pm$    0.06 &        5.03 $\pm$    0.02 &        4.25 $\pm$    0.01 &        3.24 $\pm$    0.00 \\    
SN~2006ej & host MW & 8.93 $\pm$    0.22 &        6.67 $\pm$    0.34 &        4.98 $\pm$    0.03 &        4.27 $\pm$    0.01 &        3.23 $\pm$    0.00 \\    
SN~2007S  & host MW & 8.09 $\pm$    0.54 &        6.33 $\pm$    0.48 &        4.97 $\pm$    0.04 &        4.23 $\pm$    0.01 &        3.23 $\pm$    0.00 \\    
SN~2007af & host MW & 7.66 $\pm$    0.25 &        5.96 $\pm$    0.05 &        4.95 $\pm$    0.02 &        4.20 $\pm$    0.01 &        3.18 $\pm$    0.00 \\    
SN~2007co & host MW & 8.08 $\pm$    0.59 &        6.29 $\pm$    0.40 &        4.99 $\pm$    0.04 &        4.24 $\pm$    0.01 &        3.25 $\pm$    0.00 \\    
SN~2007cq & host MW & 8.53 $\pm$    0.27 &        6.47 $\pm$    0.17 &        5.02 $\pm$    0.04 &        4.26 $\pm$    0.01 &        3.25 $\pm$    0.00 \\    
SN~2007cv & host MW & 7.02 $\pm$    0.29 &        5.88 $\pm$    0.03 &        4.96 $\pm$    0.02 &        4.20 $\pm$    0.01 &        3.18 $\pm$    0.00 \\    
SN~2007on & host MW & 8.39 $\pm$    0.09 &        6.17 $\pm$    0.05 &        4.95 $\pm$    0.02 &        4.21 $\pm$    0.01 &        3.18 $\pm$    0.00 \\    
SN~2008Q  & host MW & 7.56 $\pm$    0.13 &        6.00 $\pm$    0.02 &        4.99 $\pm$    0.02 &        4.22 $\pm$    0.01 &        3.19 $\pm$    0.00 \\    
SN~2008ec & host MW & 7.36 $\pm$    0.41 &        5.98 $\pm$    0.05 &        4.99 $\pm$    0.03 &        4.23 $\pm$    0.01 &        3.21 $\pm$    0.00 \\    
SN~2005am & host CSLMC &  9.74 $\pm$    0.39 &        7.10 $\pm$    0.06 &        4.85 $\pm$    0.06 &        3.13 $\pm$    0.01 &        1.83 $\pm$    0.00 \\    
SN~2005cf & host CSLMC &  8.96 $\pm$    0.54 &        6.97 $\pm$    0.06 &        4.79 $\pm$    0.06 &        3.11 $\pm$    0.00 &        1.83 $\pm$    0.00 \\    
SN~2005df & host CSLMC &  9.92 $\pm$    0.37 &        7.12 $\pm$    0.05 &        4.81 $\pm$    0.06 &        3.10 $\pm$    0.01 &        1.82 $\pm$    0.01 \\    
SN~2005ke & host CSLMC & 13.13 $\pm$    0.82 &        7.89 $\pm$    0.43 &        4.68 $\pm$    0.06 &        3.02 $\pm$    0.01 &        1.82 $\pm$    0.00 \\    
SN~2006dm & host CSLMC & 11.08 $\pm$    1.01 &        7.55 $\pm$    0.16 &        5.02 $\pm$    0.07 &        3.20 $\pm$    0.01 &        1.89 $\pm$    0.00 \\    
SN~2006ej & host CSLMC & 14.82 $\pm$    0.57 &        8.98 $\pm$    0.85 &        4.87 $\pm$    0.08 &        3.22 $\pm$    0.01 &        1.88 $\pm$    0.00 \\    
SN~2007S  & host CSLMC & 12.03 $\pm$    1.38 &        7.93 $\pm$    1.01 &        4.85 $\pm$    0.11 &        3.17 $\pm$    0.01 &        1.89 $\pm$    0.00 \\    
SN~2007af & host CSLMC & 11.45 $\pm$    0.65 &        7.24 $\pm$    0.11 &        4.78 $\pm$    0.06 &        3.11 $\pm$    0.01 &        1.82 $\pm$    0.00 \\    
SN~2007co & host CSLMC & 12.36 $\pm$    1.55 &        7.95 $\pm$    0.92 &        4.90 $\pm$    0.11 &        3.18 $\pm$    0.01 &        1.91 $\pm$    0.00 \\    
SN~2007cq & host CSLMC & 13.79 $\pm$    0.70 &        8.48 $\pm$    0.44 &        4.99 $\pm$    0.10 &        3.20 $\pm$    0.01 &        1.90 $\pm$    0.00 \\    
SN~2007cv & host CSLMC &  9.46 $\pm$    0.67 &        7.01 $\pm$    0.07 &        4.80 $\pm$    0.06 &        3.11 $\pm$    0.01 &        1.83 $\pm$    0.00 \\    
SN~2007on & host CSLMC & 13.47 $\pm$    0.24 &        7.98 $\pm$    0.11 &        4.84 $\pm$    0.06 &        3.13 $\pm$    0.01 &        1.83 $\pm$    0.00 \\    
SN~2008Q  & host CSLMC & 11.27 $\pm$    0.33 &        7.34 $\pm$    0.05 &        4.89 $\pm$    0.07 &        3.15 $\pm$    0.01 &        1.84 $\pm$    0.00 \\    
SN~2008ec & host CSLMC & 10.28 $\pm$    0.97 &        7.21 $\pm$    0.10 &        4.89 $\pm$    0.08 &        3.16 $\pm$    0.01 &        1.86 $\pm$    0.00 \\    
\tablenotetext{a}{The reddening column designates the component of the
  reddening and the extinction law used in determining the
  coefficients.}
\tablenotetext{b}{The extinction coefficients correspond to $R_X = A_X/E(B-V)$.}
\enddata 
\end{deluxetable} 

\begin{deluxetable}{lllllllll} 
\tabletypesize{\small}
\rotate
\tablecaption{Host Galaxies and Distances\label{table_host} } 
\tablehead{ 
\colhead{Name}    & \colhead{Host Galaxy} & \colhead{Type} & \colhead{Redshift} & \colhead{$\mu_{z}$\tablenotemark{a}} &  
 \colhead{$\mu_{I}$\tablenotemark{b}}  &  \colhead{$\mu_{\rm MLCS31}$\tablenotemark{c}}  &  \colhead{$\mu_{\rm MLCS17}$\tablenotemark{c}}  & \colhead{Ref\tablenotemark{d}} \\ 
\colhead{ } & \colhead{ }    & \colhead{ } & \colhead{$v/c$} & \colhead{(mag)} & \colhead{(mag)} & \colhead{(mag)}  & \colhead{(mag)}  &  \colhead{} 
} 
\tablewidth{0pt} 
\startdata 

SN 2005am & NGC 2811          & Sa          &     0.00789 &  32.67 $\pm$   0.23 &   \nodata          &  32.54 $\pm$   0.10 &  32.56 $\pm$   0.10 &1, 2\\
SN 2005cf & MCG-01-39-003    & S0 pec            &     0.00646 &  32.59 $\pm$   0.24 &   \nodata          &  32.58 $\pm$   0.10 &  32.58 $\pm$   0.08  & 1,3\\
SN 2005df & NGC 1559          & SBc          &     0.00435 &  31.04 $\pm$   0.40 &  30.91 $\pm$   0.31 &  31.72 $\pm$   0.20 &   \nodata          & 1,4,5\\
SN 2005ke & NGC 1371          & Sa    &     0.00488 &  30.91 $\pm$   0.43 &  31.70 $\pm$   0.19 &  32.00 $\pm$   0.08 &  31.92 $\pm$   0.05 & 1,4,6 \\
SN 2006dm & Arp 295A          & Sc               &     0.02202 &  34.70 $\pm$   0.17 &   \nodata          &  \nodata &   \nodata            &  7,2 \\
SN 2006ej & NGC 191A          & S0               &     0.02045 &  34.51 $\pm$   0.17 &   \nodata          &  34.97 $\pm$   0.10 &  34.86 $\pm$   0.11 & 7,8\\
SN 2007S  & UGC 5378          & Sb               &     0.01388 &  33.89 $\pm$   0.18 &   \nodata          &  33.84 $\pm$   0.10 &  34.22 $\pm$   0.07  & 7,7\\
SN 2007af & NGC 5584          & Sc        &     0.00546 &  32.31 $\pm$   0.26 &   \nodata          &  32.29 $\pm$   0.10 &  32.30 $\pm$   0.08  & 1,4\\
SN 2007co & MCG+05-43-016    & Sbc                &     0.02696 &  35.30 $\pm$   0.16 &   \nodata          &  35.38 $\pm$   0.10 &  35.42 $\pm$   0.08  & 9,10\\
SN 2007cq & PGC 214810        & Sbc                &     0.02500 &  35.09 $\pm$   0.16 &   \nodata          &  35.16 $\pm$   0.12 &  35.08 $\pm$   0.10  & 9,11\\
SN 2007cv & IC 2597           & E              &     0.00756 &  32.50 $\pm$   0.24 &  33.07 $\pm$   0.20 &   \nodata          &   \nodata           & 7,12,13\\
SN 2007on & NGC 1404          & E2                &     0.00649 &  32.09 $\pm$   0.28 &  31.45 $\pm$   0.19 &   \nodata          &   \nodata          & 14,14,15 \\
SN 2008Q  & NGC 524           & S02/Sa          &     0.00794 &  32.54 $\pm$   0.24 &  31.74 $\pm$   0.20 &   \nodata          &   \nodata          & 1, 16,15 \\
SN 2008ec & NGC 7469          & Sab              &     0.01632 &  34.16 $\pm$   0.18 &   \nodata          &   \nodata          &   \nodata          & 1,17 \\

\enddata 

\tablenotetext{a}{Hubble-flow distance includes a local velocity flow
  correction and calculates a distance assuming $H_0 = 72$
  km~s$^{-1}$~Mpc$^{-1}$. A dispersion of 150 km s$^{-1}$ is included
  in the 1$\sigma$ uncertainties.}
\tablenotetext{b} {Independent (from redshift and the SN properties) distances.} 
\tablenotetext{c} {Distance modulus calculated based on MLCS2k2 fitting.}
\tablenotetext{d} {The first reference is for the host type, the
  second is for the redshift measurement (obtained from NED), and the
  third is for the independent distance if given.}

\tablerefs{1, \citealp{Sandage_Tammann_1987}; 2,
  \citealp{Theureau_etal_1998}; 3, \citealp{daCosta_etal_1998}; 4,
  \citealp{Koribalski_etal_2004}; 5, TF \citep{Tully_etal_1992}; 6,
  SBF \citealp{Tonry_etal_2001} (decreased by 0.16 magnitudes as per
  \citealp{Jensen_etal_2003}); 7, \citealp{devac_etal_1991}; 8,
  \citealp{Abazajian_etal_2003}; 9: \citealp{Jarrett_etal_2000}, 10,
  \citealp{Marzke_etal_1996}; 11, \citealp{Wood-Vasey_etal_2008}; 12,
  \citealp{Bosma_Freeman_1993}; 13, \citealp{Mieske_etal_2005}, 14,
  \citealp{Graham_etal_1998}; 15, \citealp{Jensen_etal_2003}; 16,
  \citealp{Simien_Prugniel_2000}; 17, \citealp{Keel_1996} }

\end{deluxetable} 


\begin{deluxetable}{lrrrr} 
\tablecaption{UV-Optical Peak Pseudocolors\label{table_colors}   } 
\tablehead{\colhead{Color} & \colhead{Observed} & \colhead{MW} & \colhead{MW+host(MW)} & \colhead{MW+host(CSLMC)} \\  
\colhead{} & \colhead{(mag)} & \colhead{(mag)}& \colhead{(mag)} & \colhead{(mag)} }  
\tablewidth{0pt} 
\startdata 

$b-v$              &   0.15 $\pm$ 0.11 &  0.09 $\pm$ 0.12 &  -0.03 $\pm$ 0.07 &  -0.02 $\pm$ 0.09 \\
u$-b$              &  -0.28 $\pm$ 0.16 & -0.33 $\pm$ 0.15 &  -0.43 $\pm$ 0.13 &  -0.49 $\pm$ 0.14 \\
uvw1$-b$           &   1.32 $\pm$ 0.26 &  1.25 $\pm$ 0.27 &   1.13 $\pm$ 0.22 &   1.05 $\pm$ 0.21 \\
uvw1$_{rc}-b$      &   1.73 $\pm$ 0.38 &  1.65 $\pm$ 0.39 &   1.39 $\pm$ 0.25 &   1.24 $\pm$ 0.26 \\
uvm2$-b$           &   3.61 $\pm$ 0.91 &  3.39 $\pm$ 0.98 &   3.17 $\pm$ 0.97 &   3.02 $\pm$ 0.94 \\
uvm2$-$uvw1        &   2.37 $\pm$ 0.71 &  2.23 $\pm$ 0.77 &   2.09 $\pm$ 0.77 &   1.99 $\pm$ 0.76 \\
uvm2$-$uvw1$_{rc}$ &   2.00 $\pm$ 0.72 &  1.86 $\pm$ 0.77 &   1.80 $\pm$ 0.77 &   1.74 $\pm$ 0.76 \\
uvw1$-u$           &   1.59 $\pm$ 0.17 &  1.56 $\pm$ 0.18 &   1.53 $\pm$ 0.18 &   1.51 $\pm$ 0.18 \\
uvw1$_{rc}-u$      &   2.01 $\pm$ 0.32 &  1.98 $\pm$ 0.34 &   1.78 $\pm$ 0.23 &   1.68 $\pm$ 0.21 \\
 
\enddata
\tablecomments{Average colors and the root-mean square (RMS) scatter
  for the SNe with \mb\ $< 1.6$ mag without any extinction
  corrections, with only MW extinction, with MW extinction and
  host-galaxy extinction corrected with the MW extinction law, and
  with MW extinction corrected with the MW extinction law and the host
  extinction corrected with the CSLMC law. }
\end{deluxetable} 

\clearpage


\begin{deluxetable}{lrrrrrr} 
\rotate
\tabletypesize{\small} 
\tablecaption{Absolute Peak Magnitudes Corrected with the MW Extinction Law\label{table_absmagsmw}} 
\tablehead{ 
\colhead{Name} & \colhead{\mb} & \colhead{uvm2}  & \colhead{uvw1$_{rc}$} & \colhead{$u$} & \colhead{$b$}  & \colhead{$v$}  
 } 
\tablewidth{0pt} 
\startdata 

SN~2005am &   1.52 & -15.34 $\pm$   0.35 & -17.75 $\pm$   0.31 &   \nodata          & -19.25 $\pm$   0.28 & -19.29 $\pm$   0.26 \\
SN~2005cf &   1.07 & -15.65 $\pm$   0.41 & -18.20 $\pm$   0.34 & -20.16 $\pm$   0.31 & -19.84 $\pm$   0.29 & -19.66 $\pm$   0.27 \\
SN~2005df &   1.21 & -14.56 $\pm$   0.40 & -17.09 $\pm$   0.38 &   \nodata          & -18.66 $\pm$   0.36 & -18.70 $\pm$   0.34 \\
SN~2005ke &   1.77 & -13.70 $\pm$   0.37 & -14.51 $\pm$   0.37 & -16.73 $\pm$   0.26 & -17.21 $\pm$   0.24 & -17.81 $\pm$   0.22 \\
SN~2006dm &   1.54 &   \nodata          & -17.58 $\pm$   0.53 & -19.20 $\pm$   0.45 & -18.87 $\pm$   0.38 & -18.85 $\pm$   0.31 \\
SN~2006ej &   1.39 & -16.65 $\pm$   0.31 & -17.48 $\pm$   0.32 & -19.45 $\pm$   0.21 & -18.82 $\pm$   0.20 & -18.93 $\pm$   0.21 \\
SN~2007S  &   0.92 &   \nodata          & -18.32 $\pm$   0.55 & -20.32 $\pm$   0.26 & -19.80 $\pm$   0.23 & -19.81 $\pm$   0.22 \\
SN~2007af &   1.22 & -16.51 $\pm$   0.39 & -18.12 $\pm$   0.35 & -20.00 $\pm$   0.31 & -19.60 $\pm$   0.30 & -19.58 $\pm$   0.28 \\
SN~2007co &   1.09 &   \nodata          & -17.96 $\pm$   0.49 & -19.90 $\pm$   0.27 & -19.68 $\pm$   0.24 & -19.58 $\pm$   0.21 \\
SN~2007cq &   1.04 & -18.01 $\pm$   0.42 & -18.53 $\pm$   0.34 & -19.86 $\pm$   0.26 & -19.49 $\pm$   0.24 & -19.45 $\pm$   0.23 \\
SN~2007cv &   1.33 & -15.04 $\pm$   0.66 & -17.12 $\pm$   0.55 & -19.06 $\pm$   0.46 & -18.63 $\pm$   0.40 & -18.58 $\pm$   0.33 \\
SN~2007on &   1.89 & -15.79 $\pm$   0.23 & -16.94 $\pm$   0.22 & -18.61 $\pm$   0.21 & -18.34 $\pm$   0.21 & -18.42 $\pm$   0.20 \\
SN~2008Q  &   1.40 & -15.50 $\pm$   0.66 & -17.38 $\pm$   0.54 & -18.89 $\pm$   0.46 & -18.30 $\pm$   0.40 & -18.26 $\pm$   0.33 \\
SN~2008ec &   1.08 &   \nodata          & -18.03 $\pm$   0.56 & -19.75 $\pm$   0.45 & -19.28 $\pm$   0.39 & -19.20 $\pm$   0.32 \\

\enddata 
\tablecomments{These absolute magnitudes assume a MW extinction law
  for the Galactic and host-galaxy extinction.  Hubble-flow distances
  are used unless an independent distance is known (and listed in
  Table \ref{table_host}).  The extinction-corrected \mb\ value is
  given for easier identification of individual SNe in the
  absolute-magnitude plots.}
 \end{deluxetable} 

\begin{deluxetable}{lrrrrrr} 
\rotate
\tabletypesize{\small} 
\tablecaption{Absolute Peak Magnitudes corrected with the CSLMC extinction law\label{table_absmags17}} 
\tablehead{ 
\colhead{Name} & \colhead{\mb} & \colhead{uvm2}  & \colhead{uvw1$_{rc}$} & \colhead{$u$} & \colhead{$b$}  & \colhead{$v$} 
 } 
\tablewidth{0pt} 
\startdata 

SN~2005am &   1.52 & -15.02 $\pm$   0.34 & -17.47 $\pm$   0.29 &   \nodata          & -19.02 $\pm$   0.25 & -19.11 $\pm$   0.24 \\
SN~2005cf &   1.07 & -15.63 $\pm$   0.49 & -18.13 $\pm$   0.39 & -20.02 $\pm$   0.32 & -19.66 $\pm$   0.28 & -19.49 $\pm$   0.26 \\
SN~2005df &   1.21 & -14.88 $\pm$   0.89 & -17.29 $\pm$   0.67 &   \nodata          & -18.70 $\pm$   0.42 & -18.70 $\pm$   0.36 \\
SN~2005ke &   1.77 & -13.33 $\pm$   0.40 & -14.19 $\pm$   0.36 & -16.44 $\pm$   0.24 & -16.95 $\pm$   0.21 & -17.60 $\pm$   0.20 \\
SN~2006dm &   1.54 &   \nodata          & -17.65 $\pm$   0.65 & -19.20 $\pm$   0.45 & -18.82 $\pm$   0.32 & -18.78 $\pm$   0.24 \\
SN~2006ej &   1.39 & -16.55 $\pm$   0.36 & -17.38 $\pm$   0.33 & -19.35 $\pm$   0.21 & -18.73 $\pm$   0.19 & -18.85 $\pm$   0.20 \\
SN~2007S  &   0.92 &   \nodata          & -17.86 $\pm$   0.57 & -19.58 $\pm$   0.25 & -18.91 $\pm$   0.22 & -19.00 $\pm$   0.20 \\
SN~2007af &   1.22 & -16.33 $\pm$   0.48 & -17.87 $\pm$   0.37 & -19.70 $\pm$   0.31 & -19.29 $\pm$   0.28 & -19.31 $\pm$   0.27 \\
SN~2007co &   1.09 &   \nodata          & -17.79 $\pm$   0.54 & -19.59 $\pm$   0.28 & -19.29 $\pm$   0.22 & -19.22 $\pm$   0.19 \\
SN~2007cq &   1.04 & -18.00 $\pm$   0.56 & -18.46 $\pm$   0.39 & -19.75 $\pm$   0.26 & -19.36 $\pm$   0.22 & -19.33 $\pm$   0.21 \\
SN~2007cv &   1.33 & -15.38 $\pm$   0.83 & -17.27 $\pm$   0.63 & -19.04 $\pm$   0.45 & -18.48 $\pm$   0.33 & -18.39 $\pm$   0.26 \\
SN~2007on &   1.89 & -15.79 $\pm$   0.25 & -16.94 $\pm$   0.23 & -18.61 $\pm$   0.21 & -18.34 $\pm$   0.20 & -18.42 $\pm$   0.20 \\
SN~2008Q  &   1.40 & -15.57 $\pm$   0.95 & -17.41 $\pm$   0.64 & -18.88 $\pm$   0.46 & -18.28 $\pm$   0.34 & -18.24 $\pm$   0.26 \\
SN~2008ec &   1.08 &   \nodata          & -18.23 $\pm$   0.65 & -19.73 $\pm$   0.45 & -19.11 $\pm$   0.32 & -18.98 $\pm$   0.24 \\

\enddata 
\tablecomments{These absolute magnitudes assume a MW extinction law
  for the Galactic reddening and the CSLMC extinction law for the
  host-galaxy reddening.  Hubble-flow distances are used unless an
  independent distance is known (and listed in Table
  \ref{table_host}).  The extinction-corrected \mb\ value is given for
  easier identification of individual SNe in the absolute-magnitude
  plots.}
\end{deluxetable} 

\begin{deluxetable}{lllllll} 
\tablecaption{Mean Absolute Magnitudes of SNe~Ia with $0.9<\Delta m_{15}(B)<1.6$ mag \label{table_absmagsfits} } 
\tablehead{ 
\colhead{Filter} &\colhead{Ex model} & \colhead{Mean Abs Mag} & \colhead{RMS$_{\rm avg}$} & \colhead{slope\tablenotemark{a}} & 
\colhead{Mag$_{1.1}$\tablenotemark{a}} & \colhead{RMS$_{\rm fit}$} \\  
\colhead{} &\colhead{} & \colhead{(mag)} & \colhead{(mag)} & \colhead{} & \colhead{(mag)} & \colhead{(mag)}   
}  
\tablewidth{0pt} 
\startdata 

 $M_v$       & CSLMC & -18.95 &   0.38 &   0.92 & -19.08 &   0.33 \\ 
 $M_b$       & CSLMC & -18.97 &   0.40 &   1.01 & -19.12 &   0.33 \\ 
 $M_u$       & CSLMC & -19.48 &   0.36 &   1.24 & -19.63 &   0.24 \\ 
 $M_{\rm uvw1}$       & CSLMC & -17.92 &   0.36 &   0.82 & -18.05 &   0.31 \\ 
 $M_{\rm uvw1,rc}$ & CSLMC & -17.73 &   0.39 &   1.41 & -17.95 &   0.29 \\ 
 $M_{\rm uvm2}$       & CSLMC & -15.92 &   1.02 &   3.54 & -16.59 &   0.88 \\ 
 $M_v$       &    MW  & -19.16 &   0.49 &   1.60 & -19.41 &   0.37 \\ 
 $M_b$       &    MW  & -19.19 &   0.52 &   1.74 & -19.46 &   0.38 \\ 
 $M_u$       &    MW  & -19.66 &   0.49 &   2.06 & -19.90 &   0.25 \\ 
 $M_{\rm uvw1}$       &    MW  & -18.05 &   0.45 &   1.39 & -18.28 &   0.35 \\ 
 $M_{\rm uvw1,rc}$ &    MW  & -17.80 &   0.47 &   1.54 & -18.03 &   0.36 \\ 
 $M_{\rm uvm2}$        &    MW  & -15.91 &   1.10 &   3.03 & -16.45 &   1.00 \\ 

\enddata 
\tablenotetext{a}{A linear fit to the absolute magnitudes is represented by the slope and the 
intercept at \mb\ = 1.1 mag.  } 

\end{deluxetable} 


\begin{deluxetable}{lrrrrrrr} 
\tabletypesize{\small} 
\tablecaption{Colors and Red-Tail Corrections for Other Sources \label{table_redtailmore}  } 
\tablehead{ 
\colhead{Object\tablenotemark{a}}  & \colhead{uvw2$-$uvm2}   &  \colhead{uvm2$-$uvw1}  & \colhead{uvw1$-u$}  & \colhead{$u-b$} & \colhead{$b-v$} & \colhead{rc$_{w2}$\tablenotemark{b}}  & \colhead{rc$_{w1}$\tablenotemark{b}}    \\ 
 \colhead{  } & \colhead{ (mag)   } & \colhead{  (mag)  } & \colhead{  (mag)  } &  
 \colhead{ (mag)   } & \colhead{ (mag)   } & \colhead{  (mag)  } & \colhead{  (mag)  }  } 
\tablewidth{0pt} 
\startdata 

SN 1992A & -0.49 &  1.78 &  1.52 &  0.05 &  0.21 & -0.99 & -0.38 \\
SN 1994I & -0.19 &  1.32 &  1.22 &  0.85 &  0.93 & -0.90 & -0.61 \\
SN 1999em & -0.05 &  0.80 &  0.88 & -0.91 &  0.12 & -0.37 &  0.01 \\
3000 K BB  & -2.44 &  3.60 &  1.47 &  1.05 &  1.70 & (-4.85) & -1.58 \\
4000 K BB  & -0.57 &  2.03 &  1.54 &  0.36 &  1.16 & (-2.25) & -0.55 \\
5000 K BB  &  0.10 &  1.28 &  1.28 & -0.07 &  0.82 & -1.20 & -0.20 \\
6000 K BB  &  0.31 &  0.86 &  0.99 & -0.36 &  0.60 & -0.72 & -0.06 \\
7000 K BB  &  0.37 &  0.59 &  0.73 & -0.57 &  0.44 & -0.48 &  0.01 \\
8000 K BB  &  0.36 &  0.40 &  0.52 & -0.72 &  0.32 & -0.34 &  0.04 \\
9000 K BB  &  0.33 &  0.26 &  0.34 & -0.84 &  0.23 & -0.25 &  0.07 \\
10000 K BB  &  0.30 &  0.16 &  0.19 & -0.93 &  0.15 & -0.19 &  0.08 \\
15000 K BB  &  0.12 & -0.12 & -0.28 & -1.21 & -0.05 & -0.06 &  0.11 \\
20000 K BB  &  0.01 & -0.24 & -0.52 & -1.33 & -0.14 & -0.02 &  0.12 \\
30000 K BB  & -0.11 & -0.34 & -0.76 & -1.44 & -0.23 &  0.01 &  0.13 \\
a0i & -0.00 &  0.08 &  0.53 & -0.29 &  0.01 & -0.09 &  0.01 \\
a0v &  0.00 & -0.05 &  0.14 & -0.05 &  0.02 & -0.06 &  0.04 \\
b0i & -0.11 & -0.11 & -0.45 & -1.38 & -0.20 & -0.03 &  0.12 \\
b0v & -0.16 & -0.32 & -0.74 & -1.35 & -0.33 &  0.01 &  0.13 \\
f0i & -0.11 &  0.98 &  1.27 &  0.34 &  0.20 & -0.48 & -0.32 \\
f0v &  0.18 &  0.43 &  0.86 & -0.05 &  0.31 & -0.28 & -0.07 \\
g0v & -0.29 &  1.79 &  1.50 &  0.02 &  0.57 & (-1.70) & -0.32 \\
k2v & -0.40 &  1.81 &  1.49 &  0.54 &  0.91 & (-1.92) & -0.56 \\
o5v & -0.40 &  0.47 &  0.17 & -1.49 & -0.37 & -0.10 &  0.10 \\
E & -0.35 &  1.39 &  1.51 &  0.67 &  0.89 & -0.86 & -0.71 \\
Im & -0.02 & -0.05 &  0.08 & -0.54 &  0.32 & -0.06 &  0.08 \\
Sab & -0.34 &  0.90 &  1.07 &  0.36 &  0.78 & -0.32 & -0.22 \\
Sbc & -0.08 &  0.23 &  0.34 & -0.29 &  0.60 & -0.12 &  0.04 \\
Scd & -0.16 & -0.23 & -0.11 & -0.29 &  0.40 & -0.00 &  0.08 \\
SFB 15 Gyr & -0.31 &  1.60 &  1.29 &  0.54 &  0.96 & -1.30 & -0.40 \\
SFB 1 Gyr & -0.09 & -0.25 & -0.28 & -0.51 &  0.16 & -0.01 &  0.10 \\

\enddata  
\tablenotetext{a}{The first column denotes the name of the SN,
  temperature of the blackbody, stellar type, or galaxy model (age,
  and either continuous or a single burst of star formation (SFB)
  whose spectrum was used to calculate the colors.}
\tablenotetext{b}{Red tail corrections are defined as the magnitude difference 
between observed magnitudes and synthetic magnitudes in a corresponding filter with no red tail such that
rc$_{w2}$ = $ uvw2$-uvw2$_{rc}$.  Red tail corrections for which the UV portion is less
than 20\% ($rc_{w2} < -1.65$ or $rc_{w1} < -1.60$) are given in parentheses.}
\end{deluxetable} 


\begin{deluxetable}{lrrrrrr} 
\tabletypesize{\small} 
\tablecaption{Count Rate to Flux Conversion Factors for Other Sources \label{table_fluxfactors}  } 
\tablehead{ 
\colhead{Object\tablenotemark{a}} & \colhead{F$_{\rm uvw2}$\tablenotemark{b}} & \colhead{F$_{\rm uvm2}$} & \colhead{F$_{\rm uvw1}$ } & \colhead{F$_u$ } & \colhead{F$_b$ } & \colhead{F$_v$ }   \\ 
} 
\tablewidth{0pt} 
\startdata 

SN 1992A &  2.57 &  3.82 &  0.97 &  1.42 &  1.07 &  2.18 \\
SN 1994I &  1.89 &  7.97 &  1.42 &  1.79 &  1.13 &  3.10 \\
SN 1999em &  6.08 &  6.16 &  3.69 &  1.74 &  1.61 &  2.43 \\
3000 K BB  &  0.01 &  2.90 &  0.48 &  1.30 &  1.35 &  2.61 \\
4000 K BB  &  0.34 &  5.34 &  1.84 &  1.49 &  1.43 &  2.62 \\
5000 K BB  &  1.29 &  6.71 &  3.02 &  1.58 &  1.46 &  2.61 \\
6000 K BB  &  2.44 &  7.48 &  3.73 &  1.62 &  1.47 &  2.60 \\
7000 K BB  &  3.43 &  7.92 &  4.13 &  1.64 &  1.48 &  2.59 \\
8000 K BB  &  4.19 &  8.18 &  4.35 &  1.65 &  1.48 &  2.59 \\
9000 K BB  &  4.76 &  8.33 &  4.46 &  1.66 &  1.47 &  2.58 \\
10000 K BB  &  5.18 &  8.42 &  4.52 &  1.66 &  1.47 &  2.58 \\
15000 K BB  &  6.15 &  8.51 &  4.48 &  1.66 &  1.46 &  2.56 \\
20000 K BB  &  6.43 &  8.45 &  4.35 &  1.66 &  1.46 &  2.55 \\
30000 K BB  &  6.57 &  8.34 &  4.17 &  1.65 &  1.45 &  2.54 \\
o5v &  5.75 &  5.00 &  4.59 &  1.62 &  1.43 &  2.51 \\
b0i &  6.05 &  7.64 &  4.52 &  1.64 &  1.41 &  2.53 \\
b0v &  6.31 &  8.05 &  4.28 &  1.65 &  1.43 &  2.53 \\
a0i &  6.59 &  8.16 &  3.75 &  1.38 &  1.42 &  2.55 \\
a0v &  6.35 &  8.32 &  3.95 &  1.47 &  1.37 &  2.56 \\
f0i &  4.91 &  7.10 &  2.55 &  1.18 &  1.40 &  2.57 \\
f0v &  5.12 &  8.95 &  3.35 &  1.56 &  1.40 &  2.60 \\
g0v &  0.67 &  6.80 &  2.22 &  1.57 &  1.39 &  2.61 \\
k2v &  0.44 &  6.22 &  2.37 &  1.63 &  1.39 &  2.65 \\
E &  2.09 &  9.38 &  1.63 &  1.35 &  1.36 &  2.61 \\
Im &  6.01 &  8.28 &  4.15 &  1.43 &  1.41 &  2.56 \\
Sab &  4.91 &  6.93 &  2.52 &  1.37 &  1.37 &  2.62 \\
Sbc &  5.94 &  7.60 &  3.94 &  1.55 &  1.41 &  2.64 \\
Scd &  6.52 &  8.18 &  3.87 &  1.58 &  1.44 &  2.58 \\
SFB 15 Gyr &  1.31 &  7.20 &  2.23 &  1.57 &  1.39 &  2.64 \\
SFB 1 Gyr &  6.40 &  8.13 &  4.04 &  1.54 &  1.43 &  2.56 \\

\enddata  

\tablenotetext{a}{The first column denotes the name of the SN,
  temperature of the blackbody, stellar type, or galaxy model (age,
  and either continuous or a single burst of star formation (SFB))
  whose spectrum was used to calculate the colors.}

\tablenotetext{b}{The units of the flux conversion factors are $10^{16}
  {\rm erg~cm}^{-2}{\rm \AA}^{-1} {\rm counts}^{-1}$.
  Multiplying the observed count rate (in units of counts ${\rm s}^{-1}$ 
by the conversion factors gives
  the flux density (in units of $ {\rm erg~cm}^{-2} {\rm s}^{-1} {\rm \AA}^{-1}$) 
at the Vega effective wavelengths given in Table
  \ref{table_uvotfilters}. }

\end{deluxetable} 


\begin{deluxetable}{lrrrrrrrr} 
\tabletypesize{\small} 
\tablecaption{Milky Way Extinction Law Coefficients for Other Sources \label{table_rmwvalues}  } 
\tablehead{ 
\colhead{Object\tablenotemark{a}}  & \colhead{$R_{\rm MW,uvw2}$\tablenotemark{b}}  
 & \colhead{$R_{\rm MW,uvm2}$} & \colhead{$R_{\rm MW,uvw1}$ } & \colhead{$R_{MW,u}$ } & \colhead{$R_{MW,b}$ } & \colhead{$R_{MW,v}$ } & \colhead{$R_{\rm MW,w1rc}$}   & \colhead{$R_{\rm MW,w2rc}$}   \\
} 
\tablewidth{0pt} 
\startdata 
3000 K BB  &  3.70 &  6.86 &  4.32 &  4.83 &  4.01 &  3.12 &  5.92 &  8.76 \\
4000 K BB  &  5.18 &  7.61 &  5.22 &  4.89 &  4.06 &  3.14 &  6.15 &  8.88 \\
5000 K BB  &  6.44 &  7.96 &  5.78 &  4.93 &  4.10 &  3.15 &  6.34 &  8.90 \\
6000 K BB  &  7.21 &  8.17 &  6.12 &  4.95 &  4.12 &  3.15 &  6.50 &  8.87 \\
7000 K BB  &  7.64 &  8.30 &  6.35 &  4.97 &  4.13 &  3.15 &  6.62 &  8.83 \\
8000 K BB  &  7.89 &  8.40 &  6.52 &  4.98 &  4.14 &  3.16 &  6.73 &  8.79 \\
9000 K BB  &  8.05 &  8.47 &  6.64 &  4.99 &  4.15 &  3.16 &  6.81 &  8.76 \\
10000 K BB  &  8.14 &  8.53 &  6.74 &  5.00 &  4.16 &  3.16 &  6.89 &  8.72 \\
15000 K BB  &  8.31 &  8.67 &  7.03 &  5.02 &  4.18 &  3.17 &  7.12 &  8.61 \\
20000 K BB  &  8.33 &  8.73 &  7.17 &  5.03 &  4.19 &  3.17 &  7.24 &  8.55 \\
30000 K BB  &  8.33 &  8.78 &  7.30 &  5.04 &  4.20 &  3.17 &  7.35 &  8.49 \\
o5v &  7.90 &  8.22 &  6.62 &  5.03 &  4.21 &  3.17 &  6.71 &  8.23 \\
b0i &  8.23 &  8.64 &  7.06 &  5.04 &  4.20 &  3.17 &  7.12 &  8.45 \\
b0v &  8.31 &  8.78 &  7.28 &  5.03 &  4.21 &  3.17 &  7.34 &  8.46 \\
a0i &  8.19 &  8.71 &  6.66 &  4.94 &  4.17 &  3.17 &  6.97 &  8.60 \\
a0v &  8.24 &  8.72 &  6.82 &  4.96 &  4.16 &  3.17 &  7.08 &  8.58 \\
f0i &  7.25 &  8.41 &  5.73 &  4.88 &  4.14 &  3.16 &  6.50 &  8.67 \\
f0v &  7.90 &  8.64 &  6.27 &  4.95 &  4.13 &  3.16 &  6.70 &  8.83 \\
g5v &  5.89 &  7.98 &  5.45 &  4.93 &  4.08 &  3.15 &  6.14 &  9.07 \\
k2v &  5.39 &  7.76 &  5.26 &  4.93 &  4.04 &  3.14 &  6.23 &  8.85 \\
E &  6.30 &  8.51 &  5.21 &  4.87 &  4.05 &  3.14 &  6.44 &  8.78 \\
Im &  8.22 &  8.67 &  6.91 &  4.97 &  4.14 &  3.15 &  7.09 &  8.53 \\
Sab &  7.41 &  8.45 &  5.85 &  4.91 &  4.07 &  3.14 &  6.51 &  8.45 \\
Sbc &  8.04 &  8.56 &  6.61 &  4.99 &  4.11 &  3.14 &  6.84 &  8.49 \\
Scd &  8.27 &  8.77 &  7.09 &  4.97 &  4.13 &  3.15 &  7.28 &  8.47 \\
SFB 15 Gyr &  5.96 &  8.03 &  5.42 &  4.92 &  4.04 &  3.14 &  6.25 &  8.81 \\
SFB 1 Gyr &  8.30 &  8.77 &  7.13 &  4.99 &  4.15 &  3.16 &  7.27 &  8.51 \\
SN 1992A &  6.44 &  8.06 &  5.45 &  4.91 &  4.16 &  3.16 &  6.11 &  8.63 \\
SN 1994I &  6.32 &  8.30 &  5.35 &  4.88 &  4.10 &  3.14 &  6.49 &  8.70 \\
SN 1999em &  7.63 &  8.37 &  6.18 &  5.00 &  4.16 &  3.16 &  6.41 &  8.64 \\

\enddata  
\tablenotetext{a}{The first column denotes the name of the SN,
  temperature of the blackbody, stellar type, or galaxy model (age,
  and either continuous or a single burst of star formation (SFB))
  whose spectrum was used to calculate the colors.}
\tablenotetext{b}{The extinction coefficients correspond to $R_{MW,X} = A_{MW,X}/E(B-V)$.}
\end{deluxetable} 

\begin{deluxetable}{lrrrrrrrr} 
\rotate
\tabletypesize{\small} 
\tablecaption{Circumstellar LMC Extinction Law Coefficients for Other Sources \label{table_rcslmcvalues}  } 
\tablehead{ 
\colhead{Object\tablenotemark{a}}  & \colhead{$R_{\rm CSLMC,uvw2}$\tablenotemark{b}}  
 & \colhead{$R_{\rm CSLMC,uvm2}$} & \colhead{$R_{\rm CSLMC,uvw1}$ } & \colhead{$R_{CSLMC,u}$ } & \colhead{$R_{CSLMC,b}$ } & \colhead{$R_{CSLMC,v}$ } & \colhead{$R_{\rm CSLMC,w1rc}$}   & \colhead{$R_{\rm CSLMC,w2rc}$}   \\
} 
\tablewidth{0pt} 
\startdata 

SN1992A &  8.76 & 12.57 &  6.05 &  4.68 &  3.06 &  1.81 &  7.62 & 14.63 \\
SN1994I &  8.59 & 13.19 &  5.87 &  4.61 &  2.96 &  1.78 &  8.55 & 14.73 \\
SN1999em & 11.90 & 13.39 &  7.80 &  4.92 &  3.06 &  1.81 &  8.36 & 14.71 \\
3000 K BB &  2.86 &  9.43 &  3.72 &  4.47 &  2.84 &  1.77 &  7.15 & 14.44 \\
4000 K BB &  5.76 & 11.35 &  5.56 &  4.63 &  2.91 &  1.78 &  7.69 & 14.82 \\
5000 K BB &  8.65 & 12.26 &  6.83 &  4.72 &  2.96 &  1.79 &  8.15 & 14.95 \\
6000 K BB & 10.56 & 12.81 &  7.64 &  4.79 &  2.99 &  1.80 &  8.53 & 14.98 \\
7000 K BB & 11.72 & 13.18 &  8.20 &  4.83 &  3.02 &  1.80 &  8.85 & 14.96 \\
8000 K BB & 12.44 & 13.44 &  8.61 &  4.87 &  3.03 &  1.80 &  9.12 & 14.93 \\
9000 K BB & 12.90 & 13.64 &  8.92 &  4.89 &  3.05 &  1.80 &  9.35 & 14.89 \\
10000 K BB & 13.20 & 13.79 &  9.18 &  4.92 &  3.06 &  1.81 &  9.54 & 14.86 \\
15000 K BB & 13.83 & 14.19 &  9.95 &  4.98 &  3.09 &  1.81 & 10.17 & 14.74 \\
20000 K BB & 14.01 & 14.35 & 10.34 &  5.00 &  3.10 &  1.82 & 10.51 & 14.68 \\
30000 K BB & 14.13 & 14.50 & 10.70 &  5.03 &  3.12 &  1.82 & 10.84 & 14.62 \\
o5v & 13.14 & 13.06 &  8.93 &  5.03 &  3.14 &  1.82 &  9.17 & 14.23 \\
b0i & 13.85 & 14.13 & 10.05 &  5.03 &  3.11 &  1.82 & 10.22 & 14.55 \\
b0v & 14.11 & 14.49 & 10.66 &  5.02 &  3.13 &  1.82 & 10.81 & 14.59 \\
a0i & 13.53 & 14.29 &  9.03 &  4.74 &  3.08 &  1.81 &  9.79 & 14.74 \\
a0v & 13.71 & 14.33 &  9.45 &  4.81 &  3.05 &  1.81 & 10.09 & 14.71 \\
f0i & 10.92 & 13.48 &  6.73 &  4.58 &  3.03 &  1.81 &  8.58 & 14.76 \\
f0v & 12.49 & 14.08 &  8.02 &  4.79 &  3.01 &  1.80 &  9.07 & 15.01 \\
g5v &  7.31 & 12.33 &  6.07 &  4.73 &  2.93 &  1.79 &  7.67 & 15.31 \\
k2v &  6.23 & 11.75 &  5.66 &  4.74 &  2.87 &  1.78 &  7.89 & 14.80 \\
E  &  8.64 & 13.72 &  5.57 &  4.55 &  2.90 &  1.78 &  8.44 & 15.04 \\
Im & 13.74 & 14.20 &  9.66 &  4.83 &  3.03 &  1.80 & 10.11 & 14.66 \\
Sab  & 11.65 & 13.61 &  7.06 &  4.66 &  2.93 &  1.79 &  8.63 & 14.54 \\
Sbc & 13.26 & 13.92 &  8.90 &  4.89 &  2.97 &  1.79 &  9.48 & 14.58 \\
Scd & 13.98 & 14.48 & 10.17 &  4.83 &  3.01 &  1.80 & 10.65 & 14.60 \\
SFB 15 Gyr &  7.65 & 12.47 &  6.02 &  4.72 &  2.88 &  1.78 &  7.93 & 14.97 \\
SFB 1 Gyr & 14.02 & 14.46 & 10.26 &  4.90 &  3.04 &  1.80 & 10.60 & 14.64 \\

\tablenotetext{a}{The first column denotes the name of the SN,
  temperature of the blackbody, stellar type, or galaxy model (age,
  and either continuous or a single burst of star formation (SFB) )
  whose spectrum was used to calculate the colors.}
\tablenotetext{b}{The extinction coefficients correspond to $R_{CSLMC,X} = A_{CSLMC,X}/E(B-V)$.}

\enddata  
\end{deluxetable} 


\end{document}